\documentclass[sigconf]{acmart}

\usepackage{float}
\usepackage{subfigure}
\usepackage{tikz}
\usepackage{amsmath,amsfonts}
\usepackage{algorithmic}
\usepackage[ruled,linesnumbered]{algorithm2e}
\usepackage{soul}
\usepackage{multirow}
\usepackage{wrapfig}
\usepackage{dutchcal}
\usepackage{pifont}
\usepackage{adjustbox}

\usepackage{enumitem}

\usepackage{makecell}

\newcommand{\DEL}[1]{\iffalse #1 \fi}

\usepackage{filecontents}

\usepackage{cleveref}
\crefname{section}{§}{§§}
\Crefname{section}{§}{§§}
\crefformat{section}{§#2#1#3}

\usepackage{ifthenx}
\newcommand*\circled[1]{\tikz[baseline=(char.base)]{
		\node[shape=circle,draw,inner sep=0.5pt] (char) {#1};}}

\newtheorem{obs}{\textbf{Observation}}
\usepackage[most]{tcolorbox}
\definecolor{light-gray}{gray}{0.95}
\tcolorboxenvironment{obs}{
   center,
   boxsep=0.1pt,
   top=1.0pt,
   bottom=1.0pt,
   left=6pt,
   right=6pt,
    width=\columnwidth,
    colframe=light-gray,
    colback=light-gray
}

\usepackage[most]{tcolorbox}
\definecolor{light-gray}{gray}{0.95}

\newcommand{\approptoinn}[2]{\mathrel{\vcenter{
  \offinterlineskip\halign{\hfil$##$\cr
    #1\propto\cr\noalign{\kern2pt}#1\sim\cr\noalign{\kern-2pt}}}}}

\def\DONE{{\color{red}DONE }}

\usepackage{url}

\newcommand{\squishlist}{
\begin{list}{$\bullet$}
  { \setlength{\itemsep}{0pt}
     \setlength{\parsep}{0pt}
     \setlength{\topsep}{0pt}
     \setlength{\partopsep}{0pt}
     \setlength{\leftmargin}{0em}
     \setlength{\labelwidth}{0em}
     \setlength{\labelsep}{0.2em} } }

\newcommand{\squishlisttwo}{
\begin{list}{$\bullet$}
  { \setlength{\itemsep}{0pt}
     \setlength{\parsep}{0pt}
    \setlength{\topsep}{0pt}
    \setlength{\partopsep}{0pt}
    \setlength{\leftmargin}{2em}
    \setlength{\labelwidth}{1.5em}
    \setlength{\labelsep}{0.5em} } }

\newcommand{\squishend}{
  \end{list}  }

\newlength{\defaultcolumnsep}
\newcommand{\sys}{{HACK}\xspace}

\renewcommand\footnotetextcopyrightpermission[1]{} 
\setcopyright{none}

\acmDOI{}

\acmISBN{}

\acmConference[]{arXiv}
\acmYear{2025}

\acmPrice{}

\begin{document}


\title{\sys: Homomorphic Acceleration via Compression of the Key-Value Cache for Disaggregated LLM Inference}


\author{Zeyu Zhang}
\affiliation{
\institution{University of Virginia}
}
\author{Haiying Shen}
\affiliation{
\institution{University of Virginia}
}
\author{Shay Vargaftik}
\affiliation{
\institution{VMware Research}
}
\author{Ran Ben Basat}
\affiliation{
\institution{University College London}
}
\author{Michael Mitzenmacher}
\affiliation{
\institution{Harvard University}
}
\author{Minlan Yu}
\affiliation{
\institution{Harvard University}
}



\renewcommand{\shortauthors}{}

\begin{abstract}
Disaggregated Large Language Model (LLM) inference has gained popularity as it separates the computation-intensive prefill stage from the memory-intensive decode stage, avoiding the prefill-decode interference and improving resource utilization. However, transmitting Key-Value (KV) data between the two stages can be a bottleneck, especially for long prompts.
Additionally, the computation time overhead for prefill and decode is key for optimizing Job Completion Time (JCT),
and KV data size can become prohibitive for long prompts and sequences.
Existing KV quantization methods can alleviate the transmission bottleneck and reduce memory requirements, but they introduce significant dequantization overhead, exacerbating the computation time.

We propose \underline{H}omomorphic \underline{A}cceleration via \underline{C}ompression of the \underline{K}V cache (\sys) for disaggregated LLM inference. \sys eliminates the heavy KV dequantization step, and directly performs computations on quantized KV data to approximate and reduce the cost of the expensive matrix-multiplication step.
Extensive trace-driven experiments show that \sys reduces JCT by up to 70.9\% compared to disaggregated LLM inference baseline and by up to 52.3\% compared to state-of-the-art KV quantization methods. 

\end{abstract}

\maketitle

\section{Introduction}\label{sec:intro}

LLM inference is the process of generating answers to incoming requests (prompts), and its optimization is crucial for reducing costs, enhancing scalability, lowering energy consumption, and enabling deployment on commodity GPUs.
LLM inference has two stages: prefill and decode. The prefill stage processes the prompt, incurring significant compute overhead that scales with both prompt length and model size. The decode stage generates one token at a time per prompt and is memory-intensive, since it must store Key-Value (KV) data that expands with the sequence length.

Traditionally, prefill and decode run on the same GPU, causing interference: the prefill stage’s compute load delays the decode stage, whereas the decode’s memory footprint reduces the supported number of concurrent prefill operations.
Although high-end GPUs offer sufficient memory and computational power for both stages, their high cost is a limiting factor.
Disaggregated LLM inference addresses this issue by splitting the stages between different compute-focused GPUs (e.g., A10G, V100, T4, L4) for prefill, and larger-memory GPUs (e.g., A100, H100) for decode~\cite{distserve, splitwise, mooncake, memserve, strati2024dejavukvcachestreamingfast, a10g-inference, infer-without-infer}. This separation eliminates performance interference and improves overall GPU utilization, making LLM inference more efficient and cost-effective.

\begin{table*}[h]\vspace{-0in}
\centering
\footnotesize
\begin{adjustbox}{max width=\linewidth}
\begin{tabular}{|c||c|c|c|c|c|}
\hline
\multirow{2}{*}{}
& \multirow{2}{*}{\shortstack{Reducing KV \\ communication}}
& \multirow{2}{*}{\shortstack{Reducing  compute}}
& \multirow{2}{*}{\shortstack{Reducing memory access \\ latency for KV}}
& \multirow{2}{*}{\shortstack{Avoiding KV \\ dequantization}}
& \multirow{2}{*}{\shortstack{Achieving high KV \\ compression rate}} \\
& & & & & \\
\hline\hline
Baseline & $\times$ & $\times$ & $\times$ & $\checkmark$ & $\times$ \\
\hline
KV quantization & $\checkmark$ & $\times$ & $\checkmark$ & $\times$ & $\checkmark$ \\
\hline
FP4/6/8 & $\checkmark$ & Require hardware & $\checkmark$ & Require hardware & $\times$ \\
\hline
\sys & $\checkmark$ & $\checkmark$ & $\checkmark$ & $\checkmark$ & $\checkmark$ \\
\hline
\end{tabular}
\end{adjustbox}
\vspace{-0in}
\caption{Features of different methods and our system \sys.}
\vspace{-0in}
\label{tab:features}
\end{table*}


The prefill GPU must send its KV data to the decode GPU over the network. However, inexpensive GPU instances from cloud providers often lack high-speed networking. For example, AWS's A10G, V100, T4, and L4 instances cost roughly 10–20 times less than A100 instances—which typically offer 400 Gbps bandwidth—but their network speeds are limited to 10–50 Gbps or lower~\cite{aws-gpu-instances}. Similarly, Tencent Cloud’s A100 instances are configured with only 5–50 Gbps bandwidth to cut costs~\cite{tencentcloud-a100}. As a result, KV transmission between the prefill and decode instances can become a bottleneck, impairing end-to-end performance. Our measurements (\cref{sec:motivation}) show that this overhead can account for up to 42.2\% of Job Completion Time (JCT). With the growing popularity of KV data sharing across requests~\cite{cachegen, mooncake} to reduce computation time in disaggregated LLM inference, the communication bottleneck is expected to worsen.


Pipelining communication to overlap with prefill computation~\cite{splitwise} can reduce communication overhead. However, if communication time greatly exceeds prefill time, this approach loses effectiveness. Additionally, when the prefill instance lacks sufficient GPU memory across decode instances to store decode-specific data, it must temporarily transfer KV data to CPU memory~\cite{strati2024dejavukvcachestreamingfast}, rendering pipelining infeasible.

Computation can also become a bottleneck; in our measurements (\cref{sec:motivation}), prefill and decode times account for up to 45.6\% and 83.3\% of JCT, respectively. Moreover, during the decode stage, GPU memory is constrained by the large volume of cached KV data~\cite{distserve, splitwise}. Memory usage can reach up to 93.7\%. Memory access latency for KV data can consume up to 33.1\% of JCT.

These issues become more pronounced in long-prompt and long-sequence applications, such as book summarization, article writing, and coding, which have surged in popularity. In fact, commercial models like Gemini 1.5~\cite{gemini-1.5} now support a context window of up to 1M. 


\DEL{As the sequence length grows, which is 9.5\%-15.1\% of JCT for short-sequence datasets and 18.6\%-21.9\% of JCT for long-sequence datasets in our measurement (\cref{sec:motivation}).
The compute in the prefill and decode stages scales significantly with the sequence length~\cite{deepspeed-ulysses, ding2023longnet, loongserve}, leading to extended computation time.
Long sequences enlarge the size of KV data, further increasing memory access latency during decode, which is approximately 17.9\%-24.6\% of the decode time for short-sequence datasets and approximately 38.2\%-49.6\% of the decode time for long-sequence datasets in our measurement (\cref{sec:motivation}).
Therefore, as such long-sequence applications gain popularity, it becomes essential to reduce or eliminate the KV communication time, the computation time, the memory usage, and the KV dequantization overhead in disaggregated LLM inference.}

KV quantization methods, such as CacheGen~\cite{cachegen} and KVQuant~\cite{kvquant}, can alleviate the transmission bottleneck and reduce memory access latency during decoding. These approaches quantize the KV data after each iteration before storing it in the cache, then retrieve and dequantize all tokens' KV data in the next decode iteration. However, they introduce significant KV dequantization overhead.
When we employ CacheGen and KVQuant in the disaggregated LLM inference, the dequantization overhead can account for up to 37.9\% of JCT (\cref{sec:motivation}).

Lower precision Floating Point (FP) formats (e.g., FP4~\cite{fp4}, FP6~\cite{fp6}, and FP8~\cite{vllm-fp8}) can reduce KV size and accelerate computation when supported by hardware. They provide up to 73\% KV compression~\cite{open-compute}, compared to the 86\%~\cite{cachegen, kvquant} achieved by CacheGen and KVQuant through mixed-precision and 2-bit quantization. However, for FP4 and FP6, current GPUs require conversion to FP8 or FP16 for computation. On GPUs without FP8 support—such as pre-H100 architectures like the A100—FP8 must be converted to FP16, \mbox{introducing additional computational overhead.}

\DEL{To harness the advantages of KV quantization while minimizing the overhead introduced by dequantization, it is critical to avoid dequantizing KV data in each decode iteration. Previous research on quantization~\cite{thc2024li} has proposed homomorphic quantization for gradient aggregation in machine learning training, enabling direct aggregation of quantized gradient data without the need for dequantization and re-quantization. Following this idea, we can also perform attention
computation directly on the low-precision quantized KV data, applying minimal additional computation to produce an approximation of the true output, thus eliminating KV dequantization overhead and improving the speed of attention computation.
}


Ideally, arithmetic operations should be executable directly on quantized KV data, eliminating dequantization overhead and accelerating computation through smaller data elements. To this end, we propose \underline{H}omomorphic \underline{A}cceleration via \underline{C}ompression of the \underline{K}V cache (\sys) for disaggregated LLM inference. \sys addresses the KV transmission bottleneck by enabling computation on quantized data while maintaining comparable inference accuracy and reducing memory constraints. Table~\ref{tab:features} summarizes the advantages of \sys over earlier approaches.


\DEL{Previous homomorphic quantization techniques~\cite{thc2024li} for gradient aggregation in machine learning training are limited to matrix addition and cannot be applied to the matrix multiplications required in attention computation, while \sys is for this purpose. 
}


In summary, our work has the following contributions:\vspace*{-1mm}
\begin{itemize}[leftmargin=1em]
    \item We propose a homomorphic  quantization method for matrix multiplication. It conducts KV-related matrix multiplication on quantized matrices without the need for dequantizing them and then applies an approximation to transform the quantized output into an approximation of the real output.
This approach reduces KV transmission latency, computation time, memory access latency, and memory demand, all while avoiding the costly overhead of KV dequantization.
    \item We further reduce the overhead of homomorphic quantization by storing data using a small amount of memory to eliminate redundant computations and the need to update the quantized data.

    \DEL{reducing the number of recomputation operations in approximation with a minimal cost of memory. Moreover, in the decode stage, we do not immediately quantize the newly generated token. Instead, we perform computation using its original values and wait until a sufficient number of new tokens have accumulated before quantizing them all at once. This approach eliminates the requantization overhead for recently generated tokens and reduces quantization error with a minimal memory overhead.}
    \item We integrate \sys into FlashAttention-2~\cite{flashattn2} and build our system on vLLM~\cite{vllm2023kwon}. We conduct extensive experiments across various models, datasets, and GPU configurations. These experiments demonstrate that \sys can reduce JCT by up to 70.9\% and 52.3\% compared to the disaggregated LLM inference baseline and state-of-the-art KV quantization methods, respectively. We open-sourced the code of \sys~\cite{hkvq-code}.
\end{itemize}


\section{Motivation}\label{sec:motivation}
We investiage the networking, computation, and memory bottlenecks in disaggregated LLM inference and demonstrate the limitations of current KV quantization methods in addressing these issues. The default experiment settings are detailed in~\cref{sec:exp_setup}.


\subsection{Bottlenecks in Disaggregated LLM~Inference}\label{sec:kv_bottleneck}
The popularity of disaggregation has led to the adoption of cheaper GPU instances for prefill to reduce costs~\cite{distserve, splitwise, mooncake, memserve, strati2024dejavukvcachestreamingfast, a10g-inference, infer-without-infer}. However, this also exacerbates certain overheads.


\begin{figure*}[h]
    \centering
    \subfigure[Varing GPU for prefill.\label{fig:comm_bottleneck_gpus}]
    {\includegraphics[width=0.245\linewidth,height=2.5cm]{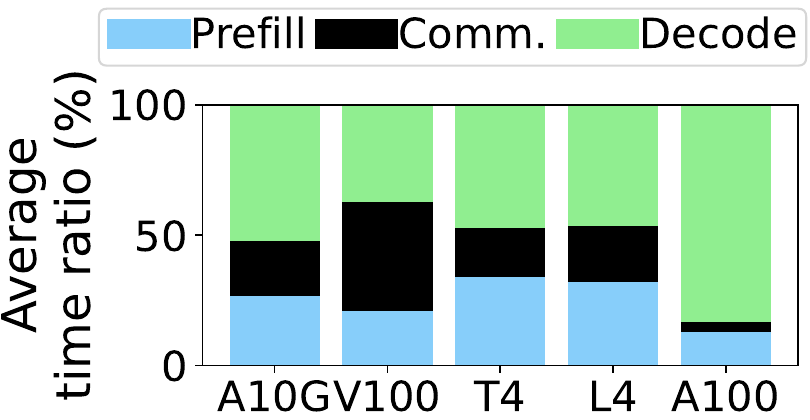}}
    \subfigure[Varying model.\label{fig:comm_bottleneck_models}]
    {\includegraphics[width=0.245\linewidth,height=2.5cm]{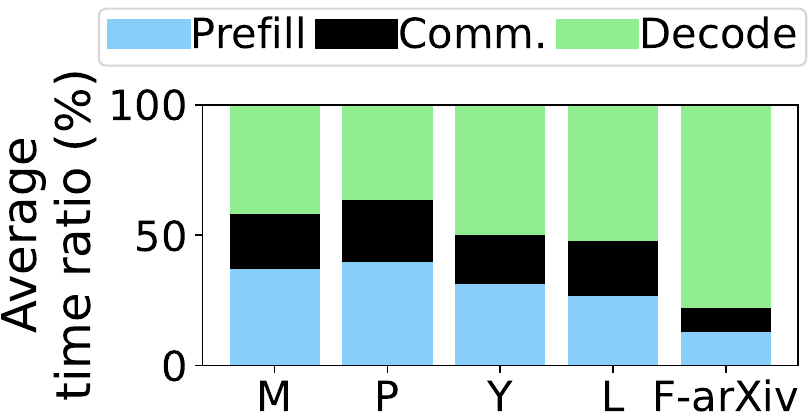}}
    \subfigure[Varying dataset.\label{fig:comm_bottleneck_datasets}]
    {\includegraphics[width=0.245\linewidth,height=2.5cm]{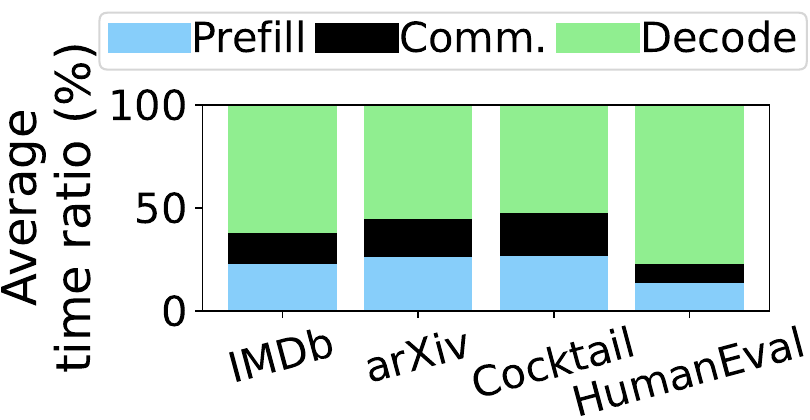}}
    \subfigure[With pipelining.\label{fig:comm_bottleneck_overlapping}]
    {\includegraphics[width=0.245\linewidth,height=2.5cm]{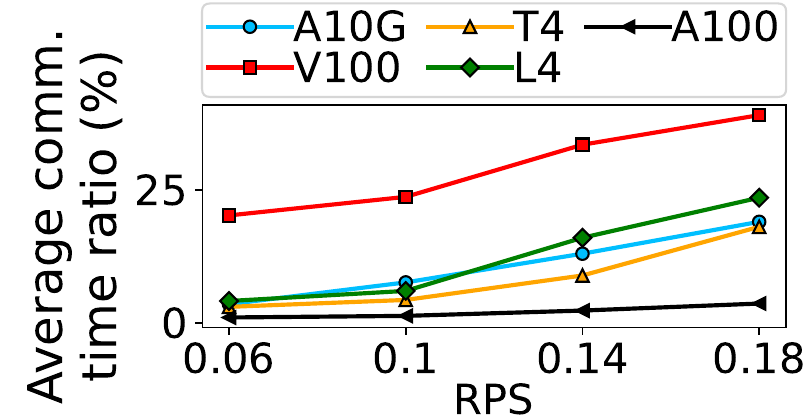}}

    \vspace{-0.2in}
    \caption{Bottlenecks in disaggregated LLM inference.}
    \label{fig:comm_bottleneck}
    \vspace{-0in}
\end{figure*}


\noindent\textbf{Network and computation overhead.}
To demonstrate the network and computation overhead when using varying prefill instances, we tested Llama-3.1 using the Cocktail on A100, T4, A10G, L4, and V100. Fig.~\ref{fig:comm_bottleneck_gpus} shows the average prefill time ratio, average communication time ratio, and average decode time ratio. The average time ratio is calculated by $\frac{1}{N}\sum_{i=1}^N(\frac{\mbox{time}_i}{\mbox{JCT}_i})$, where $\mbox{time}_i$ and $\mbox{JCT}_i$ denote the time and JCT of request $i$, respectively.
A100, equipped with a 400 Gbps network, achieves a 3.7\% average communication time ratio, while other instances with 10–50 Gbps networks range from 19.1–23.5\%.
The average prefill time ratio and the average decode time ratio are 19.7\%-41.4\% and \mbox{43.1\%-82.5\%, respectively.}

Next, we tested the performance of different models using the Cocktail dataset. Since Falcon-180B has a limitation of 2K context window, it cannot process Cocktail. Therefore, we used another long-sequence dataset, arXiv~\cite{arxiv-summarization}, for Falcon 180B, and denote it as F-arXiv.
Fig.~\ref{fig:comm_bottleneck_models} shows the average time ratios. The average communication time ratio is 11.8\% for F-arXiv and 18.7\%-25.3\% for other models.
\DEL{21\%, 25.3\%, 18.7\%, 21.9\%, and 11.8\% for models M, P, Y, L, and F-arXiv, respectively.
For the four models using Cocktail, KV communication overhead is similarly high due to the longer sequence lengths in Cocktail. F-arXiv has a smaller KV communication overhead because of the shorter sequence lengths.}
The average prefill time ratio and the average decode time ratio are 17.6\%-45.6\% and 39.8\%-81.7\%, respectively.








Fig.~\ref{fig:comm_bottleneck_datasets} presents the average time ratios for Llama-3.1 70B across various datasets on A10G prefill instances. The average communication time ratio, driven by input sequence length, varies from 9.5\% to 21.9\%. The average prefill and decode time ratios are 13.6\%-37.1\% and 54.8\%-83.3\%, respectively. The arXiv and Cocktail datasets, with long prompts and sequences, incur 15.5--43.1$\times$ higher KV communication time and 9.8--19.2$\times$ higher computation time compared to the shorter IMDb and HumanEval datasets.


\noindent\textbf{Memory overhead.}
To illustrate the memory bottleneck, we measured peak GPU memory usage on decode instances for Llama-3.1 70B across different datasets. This metric represents the ratio of memory required for parameters, KV data, and activations to the total memory capacity, ranging from 65.3\% to 93.7\% (Table~\ref{tab:peak_gpu_mem}). We also measured GPU memory access time for loading KV data during decode, with an average ratio of 16.3\%–33.1\%.


\begin{obs}~\label{obs:kv_bottleneck}
In disaggregated LLM inference, KV transmission can contribute up to 42.2\% of JCT, with prefill and decode times reaching 45.6\% and 83.3\%, GPU memory usage up to 93.7\%, and KV memory access up to 33.1\%, creating significant bottlenecks.
\DEL{In the disaggregated LLM inference, KV transmission can contribute up to 42.2\% of JCT.
The prefill time and decode time can account for up to 45.6\% and up to 83.3\% of JCT, respectively, while GPU memory usage can reach up to 93.7\%, and memory access time for KV can be up to 33.1\% of JCT. Together, these factors can become significant bottlenecks.}
\end{obs}

\DEL{\noindent\textbf{Pipelining for reducing communication overhead.}
One approach to reducing KV transmission overhead is pipelining, overlapping communication with prefill computation~\cite{splitwise}. However, pipelining has two limitations. First, it cannot fully hide communication within the prefill computation. 
In Fig.~\ref{fig:comm_bottleneck_gpus}, the communication time on V100 prefill instances is approximately twice the prefill time. Second, if the prefill instance does not pre-know the decode instance due to the insufficient GPU memory on all decode instances, it has to transfer KV data to CPU memory on the prefill instance until a decode instance has enough memory~\cite{strati2024dejavukvcachestreamingfast}, and thus pipelining cannot be realized\sh{\DONE add reference, this point also needs to be explained in Intro as we discussed.}.
\sh{\DONE is the following the third point or the second point above? if the former, change two to three above, and add Third here. If the latter, make the connection above and below tight.}
}

\noindent\textbf{Pipelining for reducing communication overhead.}
As indicated in~\cref{sec:intro}, pipelining i) cannot mitigate communication overhead when communication time significantly exceeds prefill time, and ii) is infeasible when GPU memory is insufficient on all decode instances. In case ii), the prefill instance transfers the KV data to its CPU memory~\cite{strati2024dejavukvcachestreamingfast}.
We tested Llama-3.1 with Cocktail on various prefill instances with pipelining.   Fig.~\ref{fig:comm_bottleneck_overlapping} shows the average communication time ratio with different requests-per-second (RPS).
When RPS increases from 0.06 to 0.18, the average communication time ratio on V100 rises from 21.4\% to 39.2\%, which reflects case i) above. For A10G, T4, and L4, this ratio increases from 3.3\%-4.1\% to 18.7\%-23.5\%, which reflects case ii) above.
When RPS is 0.18, the decode instances have insufficient GPU memory. On A100 with high bandwidth, this ratio grows from 1.4\% to 3.7\%.
\DEL{For V100, the pipelining cannot reduce communication overhead because of its low bandwidth.
For A10G, T4, and L4, when RPS is 0.18, the decode instances have insufficient GPU memory to serve the new requests; hence, the prefill instances store the KV data in CPU memory, making the pipelining inapplicable.}
The results demonstrate the limitations of pipelining; it is primarily suitable for scenarios with low KV transmission overhead and light workloads.
In this paper, we focus on reducing KV communication overhead through quantization, as opposed to relying on pipelining.

\subsection{Overhead for KV Quantization}
Quantization methods can be used to reduce the KV transmission overhead and the memory overhead during decode. However, they must dequantize all tokens' KV values retrieved from the KV cache before computation in every decode iteration, which incurs substantial dequantization overhead.
CacheGen~\cite{cachegen} and KVQuant~\cite{kvquant} are representative of the state-of-the-art. CacheGen focuses on leveraging KV data’s
distributional properties to encode it into more compact bitstream representations. KVQuant, on the other hand, targets low-precision KV quantization with 2-bit precision. Both can achieve up to approximately 86\% KV compression rates with approximately 98\% of baseline accuracy.

As strawman methods,
we implemented CacheGen and KVQuant in the baseline disaggregated LLM inference. On the prefill instance, the methods quantize KV data using CacheGen or KVQuant and transmit the quantized KV data to the decode instance, which dequantizes the data back to FP16 before conducting the attention computation.

%




\begin{figure}[h]
    \centering
    \subfigure[CacheGen.\label{fig:comm_bottleneck_gpus_cachegen}]
    {\includegraphics[width=0.495\linewidth,height=2.5cm]{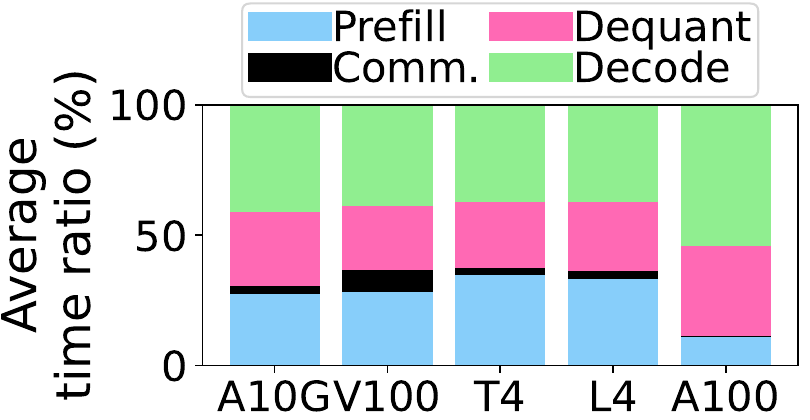}}
    \subfigure[KVQuant.\label{fig:comm_bottleneck_gpus_kvquant}]
    {\includegraphics[width=0.495\linewidth,height=2.5cm]{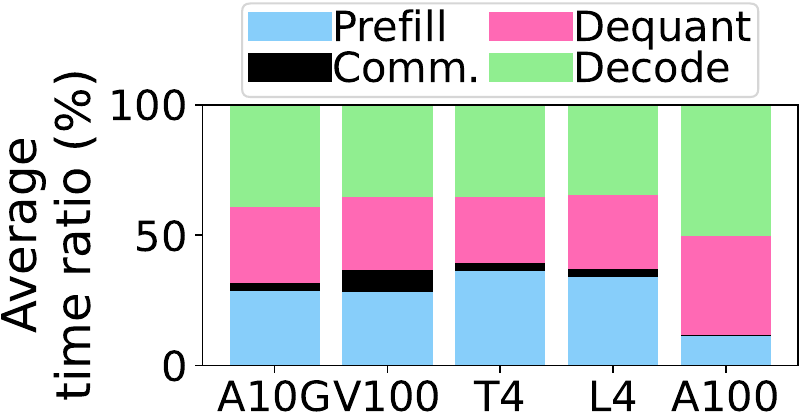}}

    \vspace{-0.2in}
    \caption{Employing KV quantization across prefill instances.}
    \label{fig:comm_bottleneck_gpus_methods}
    \vspace{-0.2in}
\end{figure}

\begin{figure}[h]
    \centering
    \subfigure[CacheGen.\label{fig:comm_bottleneck_models_cachegen}]
    {\includegraphics[width=0.495\linewidth,height=2.5cm]{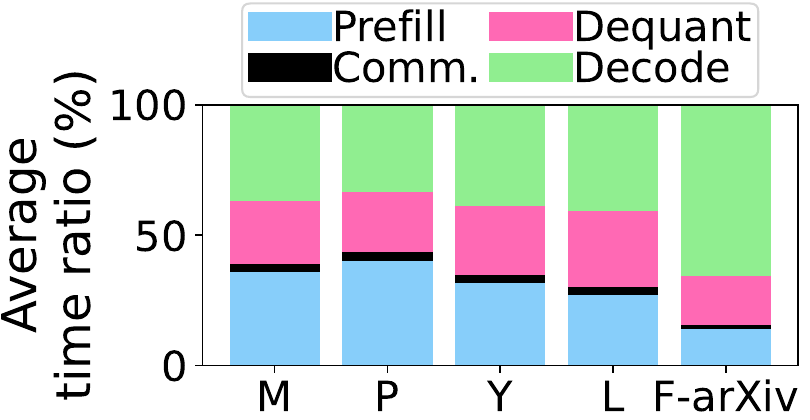}}
    \subfigure[KVQuant.\label{fig:comm_bottleneck_models_kvquant}]
    {\includegraphics[width=0.495\linewidth,height=2.5cm]{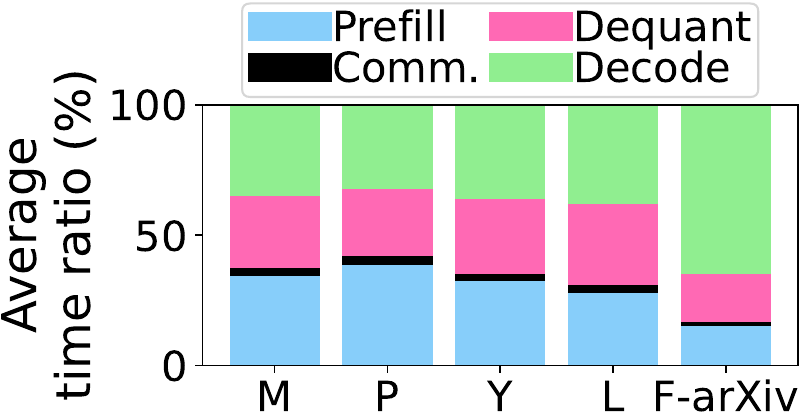}}

    \vspace{-0.2in}
    \caption{Employing KV quantization across models.}
    \label{fig:comm_bottleneck_models_methods}
    \vspace{-0in}
\end{figure}

\begin{figure}[h]
    \centering
    \subfigure[CacheGen.\label{fig:comm_bottleneck_datasets_cachegen}]
    {\includegraphics[width=0.495\linewidth,height=2.5cm]{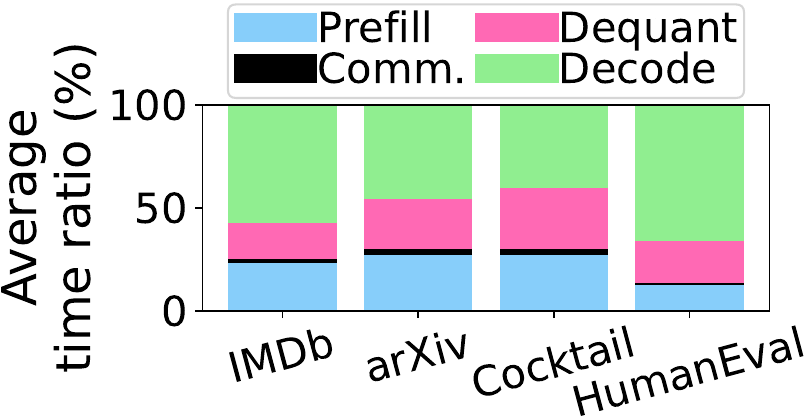}}
    \subfigure[KVQuant.\label{fig:comm_bottleneck_datasets_kvquant}]
    {\includegraphics[width=0.495\linewidth,height=2.5cm]{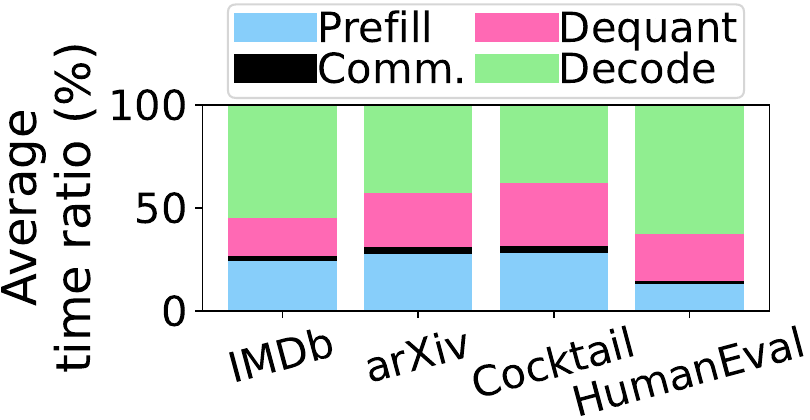}}

    \vspace{-0.2in}
    \caption{Employing KV quantization across datasets.}
    \label{fig:comm_bottleneck_datasets_methods}
    \vspace{-0.15in}
\end{figure}

Fig.~\ref{fig:comm_bottleneck_gpus_methods}, \ref{fig:comm_bottleneck_models_methods}, and~\ref{fig:comm_bottleneck_datasets_methods} illustrate the average prefill time ratio, communication time ratio, dequantization time ratio, and decode time ratio with different prefill instances, models, and datasets, respectively.
Comparing Fig.~\ref{fig:comm_bottleneck_gpus_methods} with Fig.~\ref{fig:comm_bottleneck_gpus}, we observe that
for the A100, T4, A10G, L4, and V100 prefill instances, CacheGen and KVQuant reduce the average communication time ratio
by 3.13\%-3.25\%, 16.18\%-16.39\%, 18.5\%-18.84\%, 19.81\%-20.27\%, and 33.5\% and 34.1\%, respectively. 
However, the average KV dequantization time ratios for other instances range from 26.4\% to 37.9\%.
This demonstrates that the dequantization overhead introduced in each decode iteration negatively impacts the end-to-end performance.

Comparing Fig.~\ref{fig:comm_bottleneck_models_methods} and Fig.~\ref{fig:comm_bottleneck_models}, we see that CacheGen and KVQuant reduce the average communication time ratio by 10.26\%-21.63\%.
However, for all models, they have an average dequantization time ratio of 18.2\%-30.8\%. Comparing Fig.~\ref{fig:comm_bottleneck_datasets_methods} and Fig~\ref{fig:comm_bottleneck_datasets}, we see that CacheGen and KVQuant reduce the average communication time ratio by 8.19\%-18.73\%.
However, 
they have average dequantization time ratios of
17.2\%-30.4\%.
Long-sequence datasets have 12.4-24.9$\times$ dequantization time compared to short-sequence datasets because more KV data is dequantized.



Our measurement results from Llama-3.1 70B with different datasets indicate that CacheGen and KVQuant reduce the peak memory usage on decode instances by 15.7\%-19.6\% for short-sequence datasets and by 26.9\%-33.6\% for long-sequence datasets (Table~\ref{tab:peak_gpu_mem}). They reduce the average JCT by 13.6\%-24.1\% by decreasing memory access time for KV data.

The computation times are the same between the baseline and KV quantization methods because they all perform computation on the original FP16 data.



\begin{obs}~\label{obs:dequant_overhead}
Although KV quantization methods can effectively reduce the KV transmission overhead, memory usage, and KV memory access time in disaggregated LLM inference, they introduce additional KV dequantization overhead up to 37.9\% of JCT, which is even higher for long-sequences. 
Additionally, they cannot reduce computation time.
\end{obs}


\section{Low-Precision Floating Points}\label{sec:fp8_for_kv}

Low-precision FP4/6/8 can accelerate computation when hardware supports them.
However, NVIDIA GPUs with pre-H100 architectures do not support FP8 computation.
FP4/6 needs to be converted to FP8 or FP16 for computation, leading to conversion overhead. Furthermore, FP4/6/8 cannot achieve a high KV compression rate to minimize the communication overhead and KV memory access time.

Since none of the GPUs in the experiments support FP8 computation, we conducted a simulation test to measure the communication overhead and KV memory access time for FP4/6/8 using Llama-3.1 70B, Cocktail and different prefill instances.
We converted KV data from FP4/6/8 to FP16 before attention computation, and then manually halved the time spent in matrix multiplication in attention to simulate FP8 computation.
The simulation results show that FP4, FP6, and FP8 can have an average KV communication time ratio of up to 24.3\%, 32.3\%, and 37.5\%, respectively.
The average KV memory access time ratio is 10.7\%-19.4\%.
These simulation results indicate that FP4/FP6/FP8 cannot effectively minimize the communication and memory access time overhead due to the low KV compression rate.

\DEL{
\section{Low-Precision Floating Points}\label{sec:fp8_for_kv}\sh{FP8 testing should be combined into the previous subsection. It belongs to the same category of acheGen and
KVQuant.}\ran{I don't actually know why this is in the Motivation section.}

\sh{either you move it to the top and get the same observation, or add another observation here saying: Though low-precision methods coud avoid the dequantization step and reduce computation time, their compression rate is low, so are not applicable for cheap GPUs with limited bandwidth.-make sure your results support this.}

\MM{\red{Changes marked in red below.} This should come before homomorphic.  Again, you need to make clear to the reader which are the comparison points/straw men, and which is our much better solution, in this section.}

LLM developers have explored using FP8~\cite{vllm-fp8} to reduce KV size and accelerate computation. However, adopting FP8 for KV in disaggregated LLM inference involves trade-offs between hardware requirements, accuracy, and end-to-end performance \sh{this should not be the main claim. It has the same drawbacks as CacheGen and
KVQuant, and it has lower compression ratio or not so effective in reducing data size. Additionally, you can briefly use one sentence to say it is not supported by the hardwared.}.
First, NVIDIA GPUs require at least an H100 to support FP8 computation. \ran{The term `at least an H100' is not clear. }Given the high cost of H100 GPUs, they are typically used for decode instances, leaving the prefill stage without FP8 acceleration.
\red{Second, although Table~\ref{tab:accuracy_perf} shows that FP8 achieves approximately 1\% higher accuracy compared to CacheGen, KVQuant, and our proposed system\sh{our proposed system is \sys, if yes, never mention our proposed system in this section. If no, indicate what the system is. There is no point to compare FP8 and CacheGen/ KVQuant. You just need to show what I wrote in the above.} in this paper, this comes at the expense of reduced KV compression rates and can compromise the end-to-end performance.}
CacheGen, KVQuant, and our proposed system achieve around 85\% KV compression, whereas FP8 achieves only 50\%.

\red{We implemented a strawman method using vLLM FP8 for KV~\cite{vllm-fp8} for disaggregated LLM inference and tested it using Llama-3.1 with the Cocktail dataset to compare with quantization-based methods.}
Since none of the GPUs in the experiments support FP8 computation, we converted KV data from FP8 to FP16 before attention computation. We then manually halved the time spent in matrix multiplication in attention to have simulated end-to-end performance.
The simulation results indicate that, when using V100 prefill instances, the average KV transmission time across all requests accounts for up to 37\% of the end-to-end time. This highlights that FP8 still suffers from high KV transmission overhead. Compared to CacheGen and KVQuant, FP8 reduces the end-to-end time by only 7\%-11\%, as it still has high KV transmission overhead and a longer decode time caused by higher memory access latency for KV data during the decode stage. Furthermore, in the end-to-end performance evaluation in~\cref{sec:e2e_perf_test}, our FP8 simulation results show that its end-to-end time can be up to 48\% higher than our proposed system, again due to its high KV transmission overhead and decode time.
Therefore, FP8 can effectively improve end-to-end performance while maintaining high accuracy only when the hardware supports it, and sufficient bandwidth is available\sh{you need to highlight what I wrote above. Here, the point is to show it cannot effectively reduce data size, so not suitable in our scenario (which has limited bandwidth). Do not say "...sufficient bandwidth is available".}.

}

\section{Background}\label{sec:bkg}

\noindent \textbf{LLM inference basics.}
LLM inference consists of two stages: prefill and decode. During the prefill stage, the model processes a sequence of input tokens, also referred to as the prompt. The LLM processes the input tokens to generate the first token. This first token is then used in the decode stage, where it is fed back into the model to generate the next token. The generated token serves as the input for the next decode iteration, and the process repeats iteratively to generate subsequent tokens.
LLMs consist of multiple identical layers and use the attention mechanism~\cite{self-attention} in each layer to evaluate the interdependencies between tokens in a sentence across different aspects represented by different attention heads. The input tokens, represented by the embedding matrix $E$, are transformed into a query matrix $Q^h$, a key matrix $K^h$, and a value matrix $V^h$ in each attention head $h$ by:
\vspace{-0in}
\begin{equation}\label{equ:qkv}
    Q^h=EW_Q^h \mbox{, ~} K^h=EW_K^h \mbox{, ~} V^h=EW_V^h, \vspace{0in}
\end{equation} where $W_Q^h$, $W_K^h$, and $W_V^h$ are the parameter matrices.
In prefill, the input tokens' $Q^h$, $K^h$, and $V^h$ are used for self-attention computation, and $K^h$ and $V^h$ are cached in memory for reuse in the decode stage. In decode, the input token's $K^h$ and $V^h$ are respectively combined with all previous tokens' $K^h$ and $V^h$ in the cache to form the new $K^h$ and $V^h$ that will be used in the self-attention computation.
The self-attention computes the output $O^h$ of head $h$ by: 
\vspace{-0.06in}
\begin{equation}\label{eq:attention}
    O^h=\mbox{\textit{softmax}}\bigg(\frac{Q^h(K^h)^T}{\sqrt{d_h}}\bigg)V^h=P^hV^h. \vspace{-0.05in}
\end{equation}
The \textit{softmax} function operates row-wise on the input matrix $[x_{i,j}]$ as follows:
\vspace{-0.13in}
\begin{equation}\label{eq:softmax}
        \DEL{[b_{i,j}]=\mbox{Softmax}([a_{i,j}])\mbox{, for }
        b_{i,j}=}
        \frac{exp(x_{i,j})}{\sum_{k=1}^{t_i}exp(x_{i,k})}, \vspace{-0in}
\end{equation}where $t_i$ is the index of the token on row $i$.
The outputs from all heads are concatenated together along the head dimension to form the self-attention output $O$. The model further processes $O$ through a linear transformation, a Multi-Layer Perceptron (MLP), and other operations to finally produce logits, which are used to generate an output token. Self-attention accounts for a large proportion of computational overhead in the entire process.

\noindent \textbf{Disaggregated LLM Inference.}
In disaggregated LLM inference, during the prefill stage, the prompt tokens' $Q$, $K$, and $V$ are generated and used for self-attention on the prefill instance. The prefill stage outputs the first token, which is then transferred to the decode instance along with the KV data of prompt tokens.
If no decode instance has enough memory to serve the request, the prefill instance temporarily stores the first token and KV data in its CPU memory
until a decode instance is available~\cite{strati2024dejavukvcachestreamingfast}.
On the decode instance, the first token is fed to the model as the input, and its $Q$, $K$, and $V$ are generated. The input token's $K$ and $V$ are combined with the previous tokens' $K$ and $V$. The input token's $Q$ and the combined $K$ and $V$ are used for self-attention computation.
The decode iteration outputs the next token, which will be fed into the decode model as the next input, followed by the same steps in the decode stage. This process repeats until the maximum output length is reached or the End-of-Sentence (EOS) token is output.

\section{System Design}\label{sec:design}


\subsection{Overview}



\begin{figure*}[h]
    \centering
    \includegraphics[width=1\linewidth]{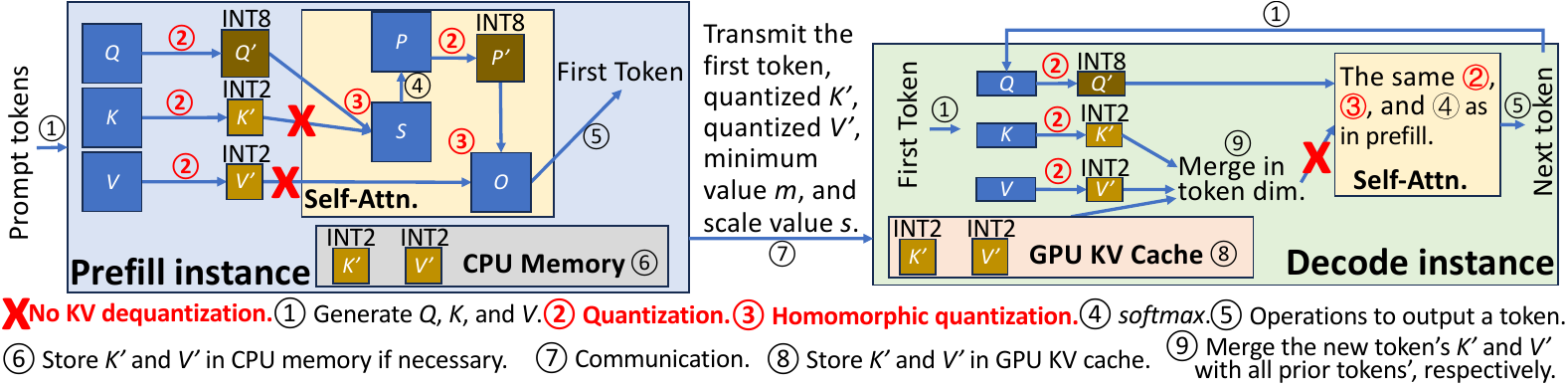}
    \vspace{-0.25in}
    \caption{Overview of \sys in disaggregated LLM inference.}
    \label{fig:overview}
    \vspace{-0in}
\end{figure*}

Observation~\ref{obs:kv_bottleneck} implies that we need to reduce the communication time, computation time, and memory access latency for KV. Therefore, we can use quantized KV values. However, this will generate high dequantization time overhead based on Observation~\ref{obs:dequant_overhead}.
To quantize KV values while eliminating the dequantization time overhead, \sys provides homomorphic quantization on KV-related matrix multiplications. Employing \sys in the disaggregated LLM inference reduces the KV data transmission time, the computation time, and the memory access latency for KV, thereby improving end-to-end response time.
Fig.~\ref{fig:overview} illustrates the workflow of employing \sys in the disaggregated LLM inference.
Since CacheGen has a KV compression rate of 86\%, and KVQuant uses 2-bit quantization to achieve a similar compression rate, we also use 2-bit quantization for KV.
The prefill instance generates $Q$, $K$, and $V$ from prompt tokens (\circled{1}) and quantizes them to $Q'$, $K'$, and $V'$ (\circled{2}).
$Q$ will be discarded right after computation, which means the quantized $Q'$ does not need to have a minimal size to save space. Thus, $Q$ uses 8-bit quantization rather than 2-bit to increase accuracy.
The first matrix multiplication between $Q'$ and $K'$ as in Eq.~\eqref{eq:attention} is performed using homomorphic quantization (\circled{3}) to output attention score $S$ without the need to dequantize $K'$, accelerated by GPU's INT8 computation capacity.
The attention score $S$ is transformed to attention probability $P$ via \textit{softmax} in Eq.~\eqref{eq:softmax} (\circled{4}), which is then quantized to $P'$ using INT8 quantization for accuracy (\circled{2}). $P'$ and $V'$ are multiplied using homomorphic quantization for acceleration (\circled{3}). The self-attention output $O$ is processed by further operations to finally output the first token (\circled{5}).
If no decode instance has enough GPU memory, the prefill instance will swap the quantized KV data to CPU memory (\circled{6}).
When a decode instance has enough memory, the prefill instance transmits the first token, $K'$, $V'$, and the quantization metadata, the minimum value $m$ and the scale value $s$, to the decode instance (\circled{7}).
$K'$ and $V'$ are stored in KV cache on the decode instance (\circled{8}).
The decode instance generates the $Q$, $K$, and $V$ from the first token (\circled{1}) and quantizes them to $Q'$, $K'$, and $V'$ (\circled{2}) as on the prefill instance. The first token's $K'$ and $V'$ are merged with all prior tokens' $K'$ and $V'$ along the token dimension, respectively (\circled{9}). The homomorphic quantization for the first token's $Q'$ and the updated $K'$ and $V'$ is performed in the same way as prefill.
The next token is generated through further operations (\circled{5}) and fed to the model to proceed to the next decode iteration (\circled{1}).

\vspace{-0in}


\subsection{Homomorphic Quantization for Matrix Multiplication}\label{sec:hoq_mm}


KV-related computation in Eq.~\eqref{eq:attention} involves two matrix multiplications: 
$Q^h(K^h)^T$ and $P^hV^h$.
We aim to perform multiplication on two quantized matrices to leverage the GPU's INT4 or INT8 computation capabilities without the need to dequantize matrix values before computation to get an approximation of the true output.
For this goal, we propose homomorphic quantization for the matrix multiplications. That is, for matrix multiplication $C=AB$, the method first quantizes $A$ and $B$ to obtain $A'$ and $B'$. It then performs the matrix multiplication $C'=A'B'$ to have the quantized output $C'$.
$C'$ is subsequently approximated into $C$ with a minimal overhead.

\begin{figure}[h]
    \centering
    \subfigure[Each row/column forms a~partition.\label{fig:hoq_matmul_1}]
    {\includegraphics[width=0.45\linewidth,height=2.3cm]{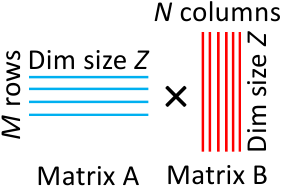}}
    \subfigure[A half row/column forms a~partition.\label{fig:hoq_matmul_2}]
    {\includegraphics[width=0.45\linewidth,height=2.3cm]{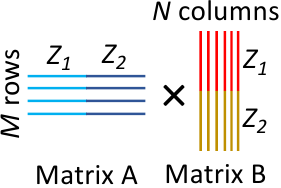}}

    \vspace{-0.2in}
    \caption{Illustration of partitioning for quantization.}
    \label{fig:hoq_matmul}
    \vspace{-0in}
\end{figure}

We use an asymmetric 2-bit stochastic quantization~\cite{squant} to reduce quantization error, which partitions the elements of each row or column into partitions.
Fig.~\ref{fig:hoq_matmul} shows examples for partitioning. We denote the dimensions of the matrices $A$ and $B$ by $M \times Z$ and $Z \times N$, respectively. Fig.~\ref{fig:hoq_matmul_1} illustrates quantization by partitioning elements in each row of $A$ and each column of $B$. The quantization partition size, denoted by $\Pi$, is the number of elements in a partition.
To increase accuracy, we can perform finer-grained quantization by setting a smaller $\Pi$, i.e., dividing the inner dimensions of the two matrices into more partitions. For example, Fig.~\ref{fig:hoq_matmul_2} illustrates quantization by partitioning elements in half of a row of $A$ and half of a column of $B$. 
In each partition $i$, the method identifies the minimum ($min_i$) and maximum ($max_i$) values of the matrix elements and computes the $scale=\frac{max_i-min_i}{2^2-1}$. Each original value $x$ in a partition is quantized to an integer $x'=round(\frac{x-min_i}{scale})$. The stochastic rounding $round(*)$ rounds $*$ to $\lfloor * \rfloor$ with probability $(\lceil * \rceil-*)/\lceil * \rceil-\lfloor * \rfloor)$ and to $\lceil * \rceil$ otherwise. 


Next, we need to estimate $C$ given $C'$. For Fig.~\ref{fig:hoq_matmul_1}, let $a_{iz}$ represent the element in the $i$-th row and $z$-th column of $A$, and $b_{zj}$ represent the element in the $z$-th row and $j$-th column of $B$. The matrix multiplication $C=AB$ can then be expressed as $c_{ij}= \sum_{z=1}^Z a_{iz}b_{zj}$ for $i\in[1,M]$ and $j\in[1,N]$.
Let $m_{a_i}$ and $s_{a_i}$ denote the minimum and scale values of $a_{iz}$. Since $a_{iz}'=round(\frac{a_{iz}-m_{a_i}}{s_{a_i}})$ and $b_{zj}'=round(\frac{b_{zj}-m_{b_j}}{s_{b_j}})$, we have $a_{iz}\approx s_{a_i}q_{a_{iz}}+m_{a_i}$ and $b_{zj}\approx s_{b_j}q_{b_{zj}}+m_{b_j}$. 
Thus, $(AB)_{ij}$ can be extended to:\vspace{-0.1in}

\DEL{\sh{\DONE how to get these equs?}
\sh{\DONE notations here that are not explained need to be explained}

\sh{\DONE you need to write high-level introduction/purpose first before giving details, e.g., for (3), you may say first: how can we conduct computation on quantized data to obtain the result similar to the result of the original data. Before you give the time complexity, also give a question like: what's the time complexity for ... operation? what is the step of this operation in the overview figure?}
}


\vspace{-0.05in}
\begin{equation}\label{eq:hoq_matmul}
\begin{split}
    \sum_z a_{iz}b_{zj}\approx\mbox{ } & s_{a_i}s_{b_j}\sum_z a_{iz}'b_{zj}' + m_{b_j}s_{a_i}\sum_z a_{iz}' + \\
    & m_{a_i}s_{b_j}\sum_z b_{zj}' + Zm_{a_i}m_{b_j},
\end{split}
\end{equation}\vspace{-0.05in}
\vspace{-0.0in}

\noindent where $\{\sum_z a_{iz}'b_{zj}',  \forall i,j\}$ is the quantized matrix multiplication that can be accelerated by INT8 computation. The other terms in Eq.~\eqref{eq:hoq_matmul} approximate $\sum_z a_{iz}'b_{zj}'$ ($C'$) into $\sum_z a_{iz}b_{zj}$ ($C$).
Eq.~\eqref{eq:hoq_matmul} is the homomorphic quantization for multiplication.


We now analyze the computational cost of matrix multiplication and approximation in Eq.~\eqref{eq:hoq_matmul}.
The computational cost for $\sum_z a_{iz}'b_{zj}'$ is $2MZN$. The remaining cost for approximating $\sum_z a_{iz}'b_{zj}'$ into $\sum_z a_{iz}b_{zj}$ is $2MN$ for multiplying $s_{a_i}$, $s_{b_i}$, and $\sum_z a_{iz}'b_{zj}'$, $MN+MZ$ for $m_{b_j}s_{a_i}\sum_z a_{iz}'$, $MN+NZ$ for $m_{a_i}s_{b_j}\sum_z b_{zj}'$, $2MN$ for $Zm_{a_i}m_{b_j}$, and $3MN$ for adding all terms, totaling $9MN+MZ+NZ$.


For Fig.~\ref{fig:hoq_matmul_2}, $AB$ can be viewed as $[A_1,A_2][B_1,B_2]^T$, which equals to $A_1B_1^T + A_2B_2^T$, where $A_1$, $A_2$, $B_1$, and $B_2$ are blocks, and each of them has multiple partitions. The multiplications $A_1B_1^T$ and $A_2B_2^T$ are performed using Eq.~\eqref{eq:hoq_matmul}, separately, followed by a summation to produce the result of $AB$.


\subsection{Homomorphic Quantization in Disaggregated LLM
Inference}\label{sec:hoqkv_for_llm}



\begin{figure}[h]
    \centering
    \includegraphics[width=0.8\columnwidth]{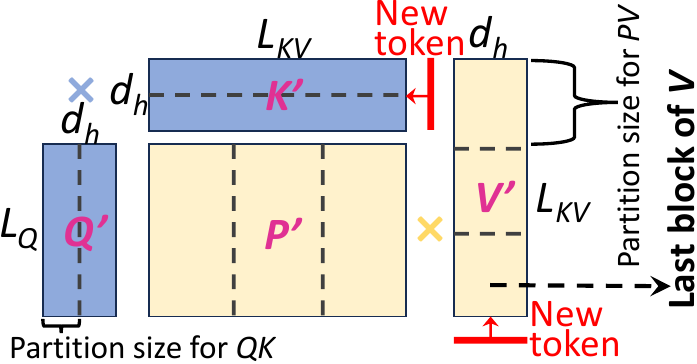}
    \vspace{-0.1in}
    \caption{Illustration of partitioning in self-attention.}
    \label{fig:hoq_self_attn}
    \vspace{-0in}
\end{figure}

We provide a detailed explanation of how our homomorphic quantization in Eq.~\eqref{eq:hoq_matmul} is applied to the self-attention in disaggregated LLM inference in Fig.~\ref{fig:overview}. Fig.~\ref{fig:hoq_self_attn} shows an example.
The inner dimension of $Q$ and $K$ is the head dimension of size $d_h$, and we divide this dimension into two partitions, so that $Q$ and $K$ both have two blocks, determined by the partition size $\Pi$. $\Pi$ must be set as a multiple of 16 to ensure efficient execution of matrix operations on the underlying GPU hardware.
$Q$ and $K$ are quantized for homomorphic quantization to output the attention score $S$, which will be processed to attention probability $P$ via $softmax$ in Eq.~\eqref{eq:softmax}.
The inner dimension of $P$ and $V$ is the sequence dimension (denoted by $L_{KV}$) rather than the head dimension. We divide this sequence dimension into three parts so that $P$ and $V$ both have three blocks, which is determined by $\Pi$.
$P$ and $V$ are quantized for homomorphic quantization to output $O$ for token generation.
During the prefill stage, the sequence lengths of $Q$ ($L_{Q}$) and $KV$ ($L_{KV}$) in Fig.~\ref{fig:hoq_self_attn} are identical, equaling to the number of prompt tokens. The prefill instance executes Eq.~\eqref{eq:attention}, during which $Q^h(K^h)^T$ and $P^hV^h$ are calculated using Eq.~\eqref{eq:hoq_matmul}. The prefill instance sends the first output token, the quantized $K'$ and $V'$, the minimum values $m$, and the maximum values $s$ to the decode instance. During the decode stage, $L_Q=1$ for the single input token. The decode instance conduct the same process as the prefill instance. After a token is generated in an iteration, it appends the token's $K'$ and $V'$ to the previous $K'$ and $V'$ along the sequence dimension $L_{KV}$ as shown in Fig.~\ref{fig:hoq_self_attn} and repeats the same process to generate the next token.


\noindent\textbf{Summation elimination.}
We first analyze the overhead of approximation in homomorphic quantization during the decode stage and then explain how we reduce this overhead.
As mentioned, in each decode iteration, matrices $Q$ and $K$ have $M=L_Q=1$, $Z=d_h$, and $N=L_{KV}$; matrices $P$ and $V$ have $M=L_Q=1$, $Z=L_{KV}$, and $N=d_h$.
The approximation in Eq.~\eqref{eq:hoq_matmul} has the computational cost of $9MN+MZ+NZ$, as explained in \cref{sec:hoq_mm}, which means the cost of the approximation is $10(d_h+L_{KV})+2d_hL_{KV}$ in each decode iteration.
If we do not use our homomorphic quantization, and we dequantize KV using the dequantization operation $sx'+m$ in every decode iteration, the cost for dequantizing KV is $2d_hL_{KV}$ for $K$ and $2d_hL_{KV}$ for $V$, totaling $4d_hL_{KV}$. Our approximation overhead is
$2d_hL_{KV}-10(d_h+L_{KV})$
less than the overhead for dequantizing KV.
In a decode iteration, since the summation term $\sum_z b_{zj}'$ for approximation in Eq.~\eqref{eq:hoq_matmul} has a cost of $NZ$, it leads to a cost of $d_hL_{KV}$ for summing the elements in quantized $K$ and a cost of $d_hL_{KV}$ for summing the elements in quantized $V$, totaling $2d_hL_{KV}$, in each decode iteration.
To reduce the summation computation, we store the sum $\sum_z b_{zj}'$ for $K$ and $V$ during decode and reuse them every iteration to avoid the recomputation cost.
For a partition size of $\Pi$ with $b$-bit integer quantization, the integer sum $\sum_z b_{zj}'$ requires at most $b+\lceil \log_2{\Pi} \rceil$ bits for storage. For example, for $\Pi=64$ with 2-bit quantization, each partition needs at most eight bits to store a sum value. Therefore, we use $b+\lceil \log_2{\Pi} \rceil$ bits to store the sum value.
This only needs a little extra memory, up to $\sim$2.7\% of the GPU memory capacity~(\cref{sec:ablation}).
Then, the final approximation cost is only $10(d_h+L_{KV})$ in each decode iteration.
The head dimension size $d_h$ is typically 128. Thus, the KV dequantization cost $4d_hL_{KV}>10(d_h+L_{KV})$ when the sequence length $L_{KV}>2.5$, and $4d_hL_{KV}$ exceeds $10(d_h+L_{KV})$ by an order of magnitude when the sequence length $L_{KV}>30$. Therefore, when $L_{KV}>30$, the approximation cost in Eq.~\eqref{eq:hoq_matmul} can be significantly reduced compared to the KV dequantization overhead in each decode iteration. The longer the sequence, the greater the reduction in the approximation cost.

\begin{wrapfigure}[11]{c}{0.25\textwidth}\vspace{-0.15in}
  \centering
  \includegraphics[width=0.25\textwidth]{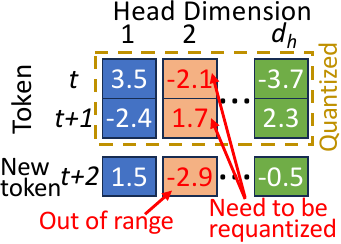}
  \vspace{-0.35in}
  \caption{An example of requantization in the last block of $V$.}
  \label{fig:v_last_group}
\end{wrapfigure}

\noindent\textbf{Requantization elimination for the last block of $V$.} After each decode iteration, 
the new token's KV values are appended to all prior tokens' KV values, as shown in Fig.~\ref{fig:hoq_self_attn}.
Since all the elements of a partition in $K$ are arranged along the fixed head dimension,
all the elements of the new token's $K$ form one or more partitions by themselves. Hence, the $[min,max]$ of the previous $K$ partitions won't be changed. In contrast, all the elements of a partition in $V$ are arranged along the sequence dimension, which is incremented by 1 after each iteration. It means the elements of the new token's $V$ will be distributed to the previous partitions.
If the number of tokens in the last block of $V$ (as shown in Fig.~\ref{fig:hoq_self_attn}) is less than the partition size $\Pi$, each element of the new token's $V$ on column $j$ is added to the existing partition on column $j$. However, the element of the new token's $V$ on column $j$ may fall outside of the previous $[min_j, max_j]$ range, leading to the necessity of updating the range and
scale $s_j$. Then, all other tokens' values on column $j$ of the last block of $V$ must be requantized. Fig.~\ref{fig:v_last_group} illustrates an example of this. When the $V$ of the new token $t+2$ is added to the last block of $V$, its value on the second column, -2.9, falls outside the previous $[\min, \max]$ range [-2.1,1.7] of that column. Consequently, the $min$ needs to be updated to -2.9, and all values on that column (-2.1 and 1.7) must be requantized based on the update $[\min, \max]$.

\setlength{\columnsep}{8pt}
To do this, we could first dequantize old quantized values using the old $min_j$ and $s_j$, quantize them and the new value by the updated $min_j$ and $s_j$, and then perform the quantized matrix multiplication for the last block of $P$ and the last block of $V$.
However, this requantization process not only increases the quantization error but also introduces extra overhead.
To address this, we do not store the quantized values of the last group of $V$ when its number of tokens does not reach the partition size $\Pi$. Instead, we store the original FP16 values for the last group of $V$ in a cache separate from the quantized KV cache, which only occupies up to 0.51\% of the GPU memory capacity~(\cref{sec:ablation}). The matrix multiplication for the last block of $P$ and the last block of $V$ are performed in FP16 format without quantization.
When the number of tokens in the last block of $V$ reaches $\Pi$, it is quantized and moved to the quantized KV cache.
As the matrix multiplication in FP16 is restricted to the last block of $V$, the associated computation time with FP16 does not scale with sequence length.
\setlength{\columnsep}{\defaultcolumnsep}

\section{Implementation}\label{sec:implementation}

\noindent\textbf{Attention backend.}
We integrated \sys with a widely adopted memory-efficient attention backend, FlashAttention-2~\cite{flashattn2}, by using OpenAI Triton~\cite{triton} and a Triton-based implementation of FlashAttention-2~\cite{triton-flashattn}. We built our system on top of vLLM~\cite{vllm2023kwon} using the modified FlashAttention-2.
Since Triton currently supports a minimum precision of INT8 for computation, we first convert the format of the quantized data from 2-bit into INT8 before performing matrix multiplication on the quantized data. This operation is performed in local GPU memory instead of global GPU memory,
to reduce the overhead for accessing data.

\noindent\textbf{Kernel fusion.} We implemented two kernels using Triton: \textit{attn\_prefill}, used during the prefill stage, and \textit{attn\_decode}, used during the decode stage. To reduce overhead, \textit{attn\_prefill} fuses the generation of QKV, QKV quantization, and self-attention with homomorphic quantization into a single kernel. In addition to these three steps, \textit{attn\_decode} integrates the process of concatenating the new token's quantized KV data with the previous tokens' KV data into the kernel to further reduce overhead. In decode, \textit{attn\_decode} separates the last block of $V$ from the quantized blocks into a buffer to enable the matrix multiplication with the original FP16 data.

\noindent\textbf{Data management.} To store the quantized KV data, along with their $m$, $s$, and sum value $\sum_z b_{zj}'$, we modified the KV cache structure of vLLM. Values $m$ and $s$ are stored in FP16.
As mentioned in~\cref{sec:hoqkv_for_llm}, we need $b+\lceil \log_2{\Pi} \rceil$ bits to store the sum $\sum_z b_{zj}'$ for a quantization partition with $b$-bit quantization. However, this may lead to memory alignment issues for a certain combination of $b$ and $|GS|$. For example, we need 9 bits to store a sum value for 2-bit quantization with a partition size of 128.
The 9-bit value can not be stored in memory at addresses that align with the GPU's natural boundaries. Therefore, we use INT16 to store a sum for this case to address the issue with memory alignment.
The memory required by the INT16 sum values only accounts for approximately 5\% of the quantized KV data.
The transmission of KV data between prefill instances and decode instances is implemented using NCCL~\cite{nccl}.

\DEL{
\noindent\textbf{Differences from TurboAttention.}
Recent work TurboAttention~\cite{turboattention} on arXiv, also performs attention computation directly on quantized KV data to accelerate attention.
However, it adopts symmetric quantization to eliminate the approximation step in Eq.~\eqref{eq:hoq_matmul}, which in turn introduces higher quantization error~\cite{turboattention}.
In contrast, our method utilizes stochastic quantization based on asymmetric quantization, resulting in lower quantization error. We store the sum value $\sum_z b_{zj}'$ for KV with minimal memory overhead, effectively reducing the approximation overhead.
Moreover, in each decode iteration, TurboAttention quantizes the last block of $V$ using INT8 before performing the quantized matrix multiplication, which introduces requantization overhead and accumulates quantization error as decoding progresses.
In contrast, \sys stores the original FP16 data for the last block of $V$ with minimal memory overhead, thereby reducing quantization error and eliminating requantization overhead.
}


\section{Performance Evaluation}

We evaluated \sys and present results in this section. The experimental setup is shown in~\cref{sec:exp_setup}. The comparison methods are the disaggregated LLM inference baseline, CacheGen and KVQaunt. 
\sys 
uses a partition size $\Pi$=64, achieving 0.16\%-0.78\% higher accuracy on average compared to CacheGen and KVQuant across all datasets and models.


\subsection{Experiment Settings}\label{sec:exp_setup}
\noindent\textbf{Testbed.}
The AWS GPU instances used in this paper are listed in Table~\ref{tab:gpu_instances}. Unless otherwise indicated, we used two p4de.24xlarge for decode~\cite{distserve, splitwise}; ten g5.12xlarge, sixteen p3.8xlarge, sixteen g4dn.12xlarge, ten g6.12xlarge, or two p4de.24xlarge for prefill so that prefill instances and decode instances have roughly similar capacities, avoiding underutilizing decode instances~\cite{splitwise}. Since existing disaggregated LLM inference systems such as DistServe~\cite{distserve} and SplitWise~\cite{splitwise} do not support Ethernet data transmission, we modified their code to enable it and integrated them onto vLLM~\cite{vllm2023kwon}
as a baseline.
The RPS was set to the maximum processing capacity, following a Poisson distribution as in~\cite{distserve}. 
The prefill and decode requests were assigned to the respective prefill and decode instances with the shortest updated queue length~\cite{splitwise}, defined by the number of queuing tokens.\looseness=-1

\begin{table}[h]
\centering
\begin{adjustbox}{max width=\columnwidth}
\begin{tabular}{ |c||c|c|c|c|c|  }
 \hline
 Name & GPUs & GPU memory & Bandwidth & vCPUs & Memory\\
 \hline
 g5.12xlarge & 4 A10G & 96 GiB & 40 Gbps & 48 & 192 GiB \\
 \hline
 p3.8xlarge & 4 V100 & 64 GiB & 10 Gbps & 32 & 244 GiB \\
 \hline
 g4dn.12xlarge & 4 T4 & 64 GiB & 50 Gbps & 48 & 192 GiB \\
 \hline
 g6.12xlarge & 4 L4 & 96 GiB & 40 Gbps & 48 & 192 GiB \\
 \hline
 p4de.24xlarge & 8 A100 & 640 GiB & 400 Gbps & 96 & 1152 GiB \\
 \hline
\end{tabular}
\end{adjustbox}
 \vspace{-0in}
\caption{GPU instances.}\label{tab:gpu_instances}
\end{table}

\begin{table}[h]\vspace{-0.1in}
\centering
\small
\begin{adjustbox}{max width=\columnwidth}
\begin{tabular}{|c||c|c|c|}
\hline
Model & A10G, L4 & V100, T4 & A100 \\
\hline
Mistral-v0.3 7B (M) & TP=4, no PP & TP=4, no PP & no TP, no PP \\
\hline
Phi-3 14B (P) & TP=2, PP=2 & TP=2, PP=2 & no TP, no PP \\
\hline
Yi 34B (Y) & TP=4, PP=2 & TP=4, PP=2 & TP=4, no PP \\
\hline
Llama-3.1 70B (L) & TP=4, PP=2 & TP=4, PP=4 & TP=4, no PP \\
\hline
Falcon 180B (F) & TP=4, PP=5 & TP=4, PP=8 & TP=4, PP=2 \\
\hline
\end{tabular}
\end{adjustbox}
\vspace{-0in}
\caption{TP and PP degrees.}
\vspace{-0in}
\label{tab:tp_pp_size}
\end{table}

\begin{table}[h]\vspace{-0in}
\centering
\small
\begin{tabular}{l|lll|lll|}
\hline
\multicolumn{1}{|l|}{Dataset} &
  \multicolumn{3}{l|}{Input length} &
  \multicolumn{3}{l|}{Output length} \\
  \cline{1-7}
  \multicolumn{1}{|l|}{} &
  \multicolumn{1}{l|}{avg} &
  \multicolumn{1}{l|}{min} &
  max &
  \multicolumn{1}{l|}{avg} &
  \multicolumn{1}{l|}{min} &
  max \\ \hline
\multicolumn{1}{|l|}{IMDb classification~\cite{imdb}}     & \multicolumn{1}{l|}{315}  & \multicolumn{1}{l|}{106}  & 821 & \multicolumn{1}{l|}{37}  & \multicolumn{1}{l|}{16} & 87  \\ \hline
\multicolumn{1}{|l|}{arXiv summarization~\cite{arxiv-summarization}}   & \multicolumn{1}{l|}{6.3K}  & \multicolumn{1}{l|}{1.6K} & 14.1K  & \multicolumn{1}{l|}{243} & \multicolumn{1}{l|}{29} & 464   \\ \hline
\multicolumn{1}{|l|}{Cocktail for IR~\cite{cocktailforir}}   & \multicolumn{1}{l|}{16.2K}  & \multicolumn{1}{l|}{9.4K} & 28.8K  & \multicolumn{1}{l|}{159} & \multicolumn{1}{l|}{44} & 246   \\ \hline
\multicolumn{1}{|l|}{HumanEval~\cite{humaneval}}   & \multicolumn{1}{l|}{204}  & \multicolumn{1}{l|}{75} & 697  & \multicolumn{1}{l|}{139} & \multicolumn{1}{l|}{11} & 552   \\ \hline
\end{tabular}%
\vspace{-0in}
\caption{Dataset properties.}
\label{tab:dataset}
\end{table}
\vspace{-0in}

\noindent\textbf{Models and datasets.}
We used a range of state-of-the-art models in the paper: Mistral AI Mistral-v0.3 7B~\cite{mistral-v0.3}, Microsoft Phi-3 14B~\cite{phi-3}, 01-ai Yi 34B~\cite{yi-model}, Meta Llama-3.1 70B~\cite{llama3.1}, and TII Falcon 180B~\cite{falcon}. We use the initial letters M, P, Y, L, and F to represent each model, respectively, for the rest of the paper. Table~\ref{tab:tp_pp_size} shows the Tensor Parallelism (TP) and Pipeline Parallelism (PP) degree of each model in different GPU instances, following practical configurations to ensure sufficient GPU memory to handle requests~\cite{distserve, splitwise}.
The datasets we use are listed in Table~\ref{tab:dataset}.
IMDb includes 27 genres of movies, TV shows, etc., collected on the Internet. It is operated by IMDb.com, Inc., a subsidiary of Amazon. The arXiv summarization has a collection of scientific publications and their summaries on arXiv.org~\cite{arxiv}. Information Retrieval (IR) is the process of retrieving relevant content from vast amounts of information based on a user query. Cocktail is a benchmark for IR, including 8 different IR tasks such as question answering, fact checking, etc. HumanEval evaluates the performance of code completion, including 164 programming problems. We use \textit{ROUGE-1}~\cite{rouge-score} and \textit{Edit Similarity (normalized Levenshtein distance)}~\cite{zhang2024hierarchicalcontextpruningoptimizing, string-similarity} as the accuracy metric for arXiv summarization and HumanEval, respectively. Unless otherwise specified, we default to testing with the Llama-3.1 70B model and the Cocktail dataset with long sequences on A10G prefill instances, which have the same architecture as the A100.

\subsection{End-to-End Time Performance}\label{sec:e2e_perf_test}

\begin{figure}[h]
    \centering
    \includegraphics[width=0.99\columnwidth]{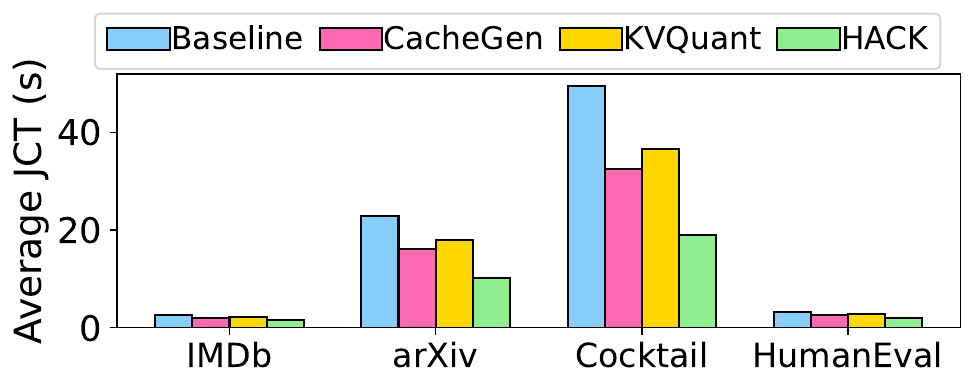}
    \vspace{-0.2in}
    \caption{Average JCT across requests for Llama-3.1 70B with varying datasets.}
    \label{fig:e2e_diff_datasets}
    \vspace{-0in}
\end{figure}

We tested Llama-3.1 70B on A10 prefill instances across different datasets. Fig.~\ref{fig:e2e_diff_datasets} shows the average JCT across all requests.
For IMDb, HumanEval, arXiv, and Cocktail, \sys reduces the average JCT by 19.2\%, 22.5\%, 36.8\%, and 41.5\%, respectively, compared to CacheGen, and by 21.2\%, 25.1\%, 40.8\%, and 45.1\%, respectively, compared to KVQuant. This is because \sys leverages quantized matrix multiplication, accelerating both prefill and decode stages while eliminating KV dequantization overhead. For IMDb, HumanEval, arXiv, and Cocktail, \sys reduces the average JCT by 38.6\%, 40.1\%, 55.3\%, and 61.6\%, respectively, compared to the baseline. This improvement is due to \sys not only accelerating prefill and decode stages but also reducing KV transmission overhead. The improvement in JCT achieved by \sys for arXiv and Cocktail is higher than IMDb and HumanEval because arXiv and Cocktail have longer sequences.

\begin{figure}[h]
    \centering
    \includegraphics[width=0.99\columnwidth]{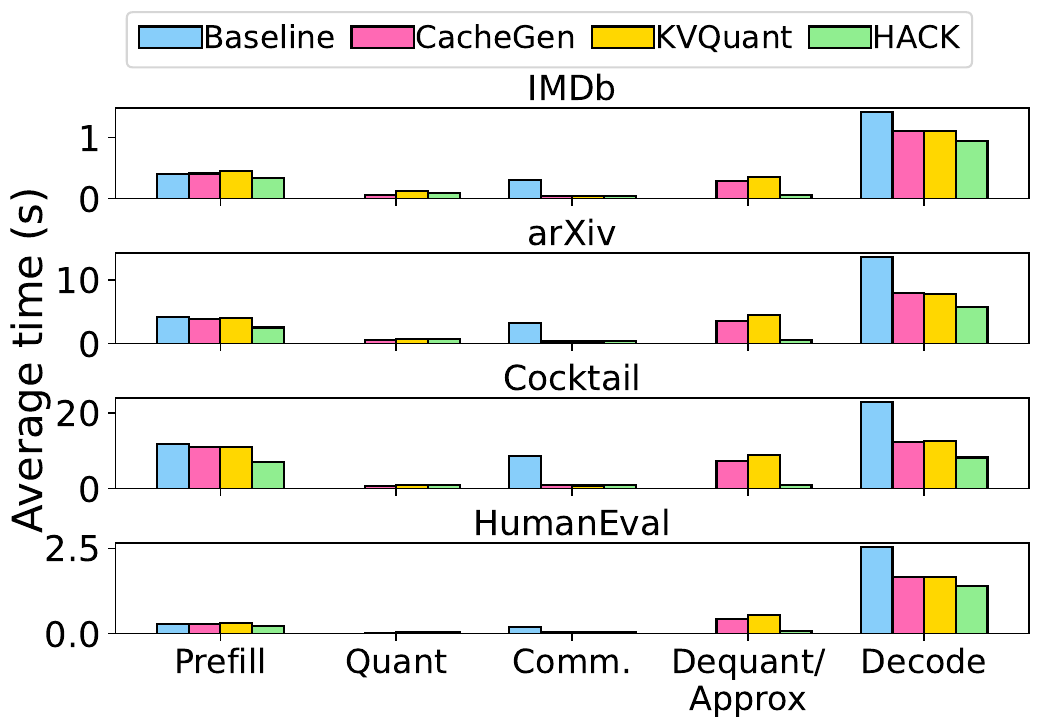}
    \vspace{-0.15in}
    \caption{Average JCT decomposition for Llama-3.1 70B with varying datasets.}
    \label{fig:e2e_decompose}
\end{figure}

We decompose the average JCT in Fig.~\ref{fig:e2e_diff_datasets} into prefill, quantization, KV communication, dequantization/approximation, and decode time, as shown in Fig.~\ref{fig:e2e_decompose}. The approximation time in homomorphic quantization is only for \sys, and the dequantization time is for other methods.
For IMDb, HumanEval, arXiv, and Cocktail, \sys achieves lower prefill times than other methods by 14.6\%-23.7\%, 17.5\%-23.4\%, 34.2\%-37.5\%, and 35.9\%-41.9\%, respectively. This is due to the acceleration of prefill computation via quantized matrix multiplication in Eq.~\eqref{eq:hoq_matmul}.
The longer the sequence, the greater the improvement in prefill time.
The quantization overhead for CacheGen, KVQuant, and \sys accounts for only 1.25\%-2.91\% of JCT, as the quantization of KV data for each token occurs only once during the entire end-to-end process. By compressing KV data to approximately 15\% of its original size, \sys, CacheGen, and KVQuant reduce KV transmission time by 80.6\%-85.4\% compared to the baseline, enabling KV transmission to account for only 1.31\%-5.4\% of JCT.
\sys eliminates the KV dequantization overhead, which accounts for 17.2\%-30.4\% of JCT in CacheGen and KVQuant, replacing it with only 1.53\%-3.18\% overhead for approximation in homomorphic quantization.

For IMDb, HumanEval, arXiv, and Cocktail, \sys reduces the decode time by 11.5\%-12.2\%, 13.7\%-14.5\%, 24.3\%-25.6\%, and 32.1\%-33.7\% compared to CacheGen and KVQuant, due to accelerated computation from quantized matrix multiplication. For short-sequence datasets IMDb and HumanEval, the improvement in decode time is only 11.5\%-14.5\% because the number of tokens in the last block of $V$ has a high proportion in the sequence length, which leads to compromised efficiency with FP16 computation. The improvement in decode time for long-sequence datasets arXiv, and Cocktail is around 20\% higher than IMDb and HumanEval because the proportion of the number of tokens of the last block of $V$ in the sequence length is small, and thus, the quantized matrix multiplication benefits the decode time more.
CacheGen and KVQuant reduce decode time by 16.5\%-38.1\% compared to the baseline because the reduced KV size improves the memory access latency. \sys also benefits from the improvement in memory access latency.
\sys is effective in reducing JCT for different datasets.

Table~\ref{tab:peak_gpu_mem} lists the peak GPU memory usage on decode instances for different datasets. CacheGen, KVQuant, and \sys can reduce the peak GPU memory usage by 13.9\%-19.6\% for short-sequence datasets IMDb and HumanEval and by 25.0\%-33.6\% for long-sequence datasets.
\sys has 0.6\% and 2.9\% higher peak GPU memory usage because it stores the sum value for the approximation step in homomorphic quantization and stores the FP16 data for the last block of $V$.

\begin{table}[h]
\centering
\begin{adjustbox}{max width=\columnwidth}
\begin{tabular}{ |c||c|c|c|c|  }
 \hline
 & IMDb & arXiv & Cocktail & HumanEval \\
 \hline
 Baseline & 65.3\% & 83.1\% & 93.7\% & 68.9\% \\
 \hline
 CacheGen & 49.6\% & 56.2\% & 61.3\% & 50.8\% \\
 \hline
 KVQuant & 48.5\% & 55.9\% & 60.1\% & 49.3\% \\
 \hline
 \sys & 51.4\% & 58.1\% & 63.0\% & 51.4\% \\
 \hline
\end{tabular}
\end{adjustbox}
 \vspace{-0in}
\caption{Peak GPU memory usage on decode instances with varying datasets.}\label{tab:peak_gpu_mem}
\end{table}

\begin{table*}[h]\vspace{-0in}
\centering
\begin{adjustbox}{max width=\linewidth}
\begin{tabular}{|c||c|c|c|c|c||c|c|c|c|c||c|c|c|c||c|c|c|c|c|}
\hline
& \multicolumn{5}{|c||}{IMDb} & \multicolumn{5}{|c||}{arXiv} & \multicolumn{4}{|c||}{Cocktail} & \multicolumn{5}{|c|}{HumanEval} \\
\hline
& M & P & Y & L & F & M & P & Y & L & F & M & P & Y & L & M & P & Y & L & F \\
\hline
Baseline & 84.81\% & 87.84\% & 93.87\% & 95.73\% & 85.63\% & 79.40\% & 86.35\% & 87.75\% & 83.79\% & 79.42\% & 75.18\% & 83.92\% & 85.25\% & 86.39\% & 89.37\% & 91.62\% & 90.79\% & 92.45\% & 85.21\% \\
\hline
\sys ($\Pi$=32) & 83.79\% & 87.08\% & 92.82\% & 94.73\% & 84.96\% & 78.53\% & 85.51\% & 86.69\% & 82.63\% & 78.75\% & 74.63\% & 83.05\% & 84.39\% & 85.54\% & 88.68\% & 90.50\% & 90.02\% & 91.28\% & 84.26\% \\
\hline
\sys ($\Pi$=64) & 83.46\% & 86.64\% & 92.31\% & 94.17\% & 84.72\% & 78.26\% & 85.08\% & 86.39\% & 82.31\% & 78.28\% & 74.42\% & 82.66\% & 84.04\% & 84.87\% & 88.19\% & 90.16\% & 89.63\% & 91.04\% & 83.82\% \\
\hline
CacheGen & 83.04\% & 86.15\% & 91.79\% & 93.85\% & 83.94\% & 77.87\% & 84.62\% & 86.02\% & 82.14\% & 77.85\% & 73.74\% & 82.24\% & 83.51\% & 84.71\% & 87.73\% & 89.75\% & 88.94\% & 90.49\% & 83.47\% \\
\hline
KVQuant & 82.97\% & 86.05\% & 91.54\% & 93.89\% & 84.12\% & 77.83\% & 84.65\% & 86.11\% & 82.08\% & 77.76\% & 73.72\% & 82.26\% & 83.42\% & 84.68\% & 87.64\% & 89.65\% & 88.93\% & 90.43\% & 83.29\% \\
\hline
\sys ($\Pi$=128) & 82.71\% & 86.07\% & 91.35\% & 93.76\% & 84.03\% & 77.72\% & 84.49\% & 86.15\% & 81.97\% & 77.94\% & 73.81\% & 82.19\% & 82.86\% & 84.65\% & 87.60\% & 89.54\% & 88.91\% & 89.77\% & 83.19\% \\
\hline
\end{tabular}
\end{adjustbox}
\vspace{-0in}
\caption{Accuracy performance.
}
\vspace{0.2in}
\label{tab:accuracy_perf}
\end{table*}

\begin{figure}[h]
    \centering
    \includegraphics[width=0.99\columnwidth]{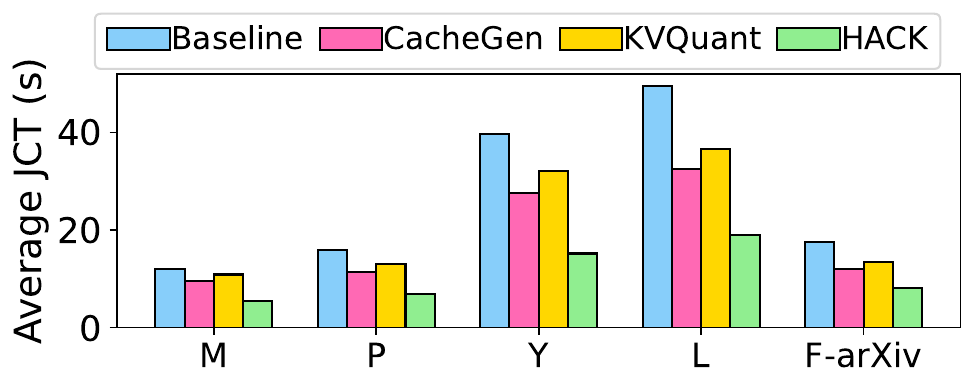}
    \vspace{-0.2in}
    \caption{Average JCT across requests for different models with Cocktail or arXiv.}
    \label{fig:e2e_diff_models}
\end{figure}

We also tested different models with the Cocktail using A10G prefill instances.
Fig.~\ref{fig:e2e_diff_models} shows the average JCT across all requests.
For M, P, Y, L, and F-arXiv, \sys reduces the average JCT by 42.4\%, 39.1\%, 44.8\%, 41.5\%, and 31.7\%, respectively, compared to CacheGen. \sys reduces the average JCT by 48.3\%, 46.5\%, 50.7\%, 45.1\%, and 37.6\%, respectively, compared to KVQuant. Compared to the baseline, \sys reduces it by 54.6\%, 57.2\%, 58.7\%, 61.6\%, and 53.3\%, respectively. The reasons are the same as those discussed in Fig.~\ref{fig:e2e_diff_datasets}.
\sys's improvement on F-arXiv is smaller than that for other models because the sequence length used in F-arXiv is shorter, capped at 2K.
\sys is effective in reducing JCT for different models.

\begin{figure}[h]
    \centering
    \includegraphics[width=0.99\columnwidth]{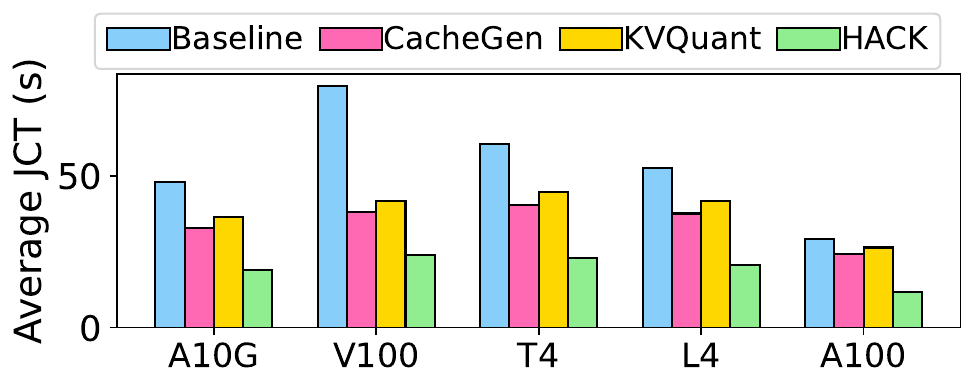}
    \vspace{-0.2in}
    \caption{Average JCT across requests for Llama-3.1 70B with Cocktail using varying prefill instances.}
    \label{fig:e2e_diff_gpus}
\end{figure}

We tested Llama-3.1 70B with the Cocktail on different prefill instances, and Fig.~\ref{fig:e2e_diff_gpus} shows the average JCT across all requests under different prefill
instances. For A10G, V100, T4, L4, and A100, \sys reduces the average JCT compared to CacheGen by 41.5\%, 37.4\%, 43.1\%, 45.3\%, and 48.5\%; compared to KVQuant by 45.1\%, 41.7\%, 46.6\%, 50.5\%, and 52.3\%; and compared to the baseline by 61.6\%, 70.9\%, 62.1\%, 59.3\%, and 60.5\%, respectively.
The improvement of \sys over CacheGen and KVQuant on V100 is the smallest because the V100 tensor core does not support INT8 matrix multiplication, making it unable to accelerate prefill computation. However, \sys achieves the highest improvement over the baseline on V100, as V100 has the lowest bandwidth, and \sys significantly improves the KV transmission speed.
\sys is effective in reducing JCT for various prefill instances with various bandwidths and compute capacity.

\subsection{Accuracy Performance}\label{sec:acc_perf}
We measured the accuracy of each method across different datasets and models, as shown in Table~\ref{tab:accuracy_perf}.
\sys uses three partition sizes $\Pi$=128, 64, and 32, denoted by \sys ($\Pi$=128), \sys ($\Pi$=64), and \sys ($\Pi$=32).
Compared to the baseline, \sys ($\Pi$=$32$), \sys ($\Pi$=$64$), CacheGen, KVQuant, and \sys ($\Pi$=$128$) exhibit accuracy losses of 0.55\%-1.17\%, 0.76\%-1.56\%, 1.44\%-2.08\%, 1.46\%-2.33\%, and 1.37\%-2.68\%, respectively.
\sys ($\Pi$=$32$) and \sys ($\Pi$=$64$) achieve higher accuracy than CacheGen and KVQuant due to the fine-grained partition size.
Meanwhile, \sys ($\Pi$=$128$) has a 0.21\%-1.27\% lower accuracy than \sys ($\Pi$=$128$) due to the larger partition size and has a slightly lower accuracy than KVQuant on average, by no more than 0.12\%.
Across different models and datasets, all methods exhibit similar accuracy loss trends.
Since the partition size of 64 is enough for \sys to have better accuracy than CacheGen and KVQuant, we use $\Pi$=$64$ as the default partition size for \sys in the \mbox{evaluation.}



\subsection{Ablation Study}\label{sec:ablation}

\begin{figure}[h]
    \centering
    \includegraphics[width=0.99\columnwidth]{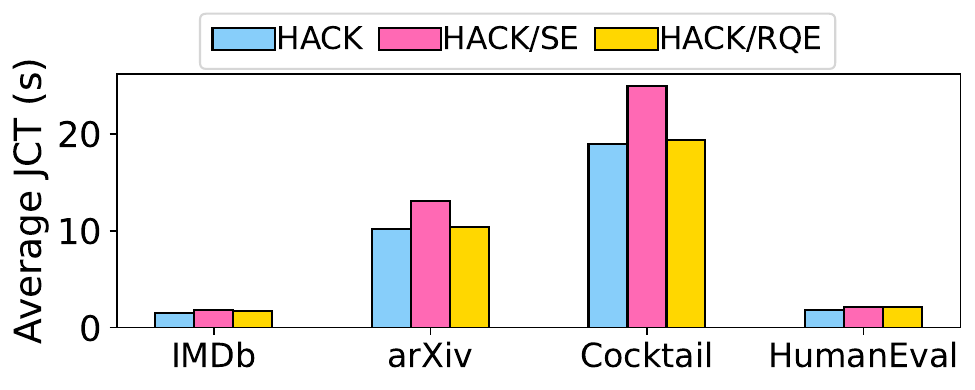}
    \vspace{-0.2in}
    \caption{Average JCT across requests for individual methods with Llama-3.1 70B.}
    \label{fig:e2e_ablation}
\end{figure}

We conducted ablation study to understand the impact of two of our optimizations on end-to-end performance and accuracy: the summation elimination for Eq.~\eqref{eq:hoq_matmul} and requantization elimination for the last block of $V$. We tested the following variations. \sys/SE represents \sys without \textit{Summation Elimination}, which means it does not store the term $\sum_z b_{zj}$ in Eq.~\eqref{eq:hoq_matmul} and recomputes it every decode iteration. \sys/RQE represents the system without \textit{ReQuantization Elimination} for the last block of $V$,
and instead checks and requantizes the last block of $V$ every decode iteration.
Fig.~\ref{fig:e2e_ablation} shows the average JCT across all requests for \sys, \sys/SE, and \sys/RQE when varying datasets.
\sys/SE has 13.8\%-15.3\% higher average JCT for the short-sequence datasets (IMDb and HumanEval) and has 22.1\%-25.9\% higher average JCT for the long-sequence datasets (arXiv and Cocktail) compared to \sys. Long sequences increase the overhead in recomputing the sum value $\sum_z b_{zj}$ for KV.
\sys/RQE has 17.8\%-21.7\% higher average JCT for the short-sequence datasets and has 0.09\%-1.2\% higher average JCT for the long-sequence datasets compared to \sys. As the sequence length increases, the proportion of tokens in the last block of $V$ compared to the total number of tokens decreases, making the overhead of requantizing the last block of $V$ comparatively smaller.
Table~\ref{tab:acc_ablation} lists the decrease in accuracy for \sys/RQE compared to \sys under different datasets. Since the quantization error caused by requantization of the last block of $V$ is only accumulated during the decode stage, the accuracy decrease is affected by the output length. IMDb has an average output length of 37, while others have an average output length of 139-243. Thus, IMDb has a 0.08\%-0.15\% less decrease in accuracy than other datasets.
SE is effective in reducing the summation overhead, and RQE can avoid accuracy loss without compromising \mbox{the end-to-end~performance.}

\begin{table}[h]\vspace{-0in}
\centering
\begin{adjustbox}{max width=\columnwidth}
\begin{tabular}{|c|c|c|c|}
\hline
IMDb & arXiv & Cocktail & HumanEval \\
\hline
-0.14\% & -0.29\% & -0.22\% & -0.25\%\\
\hline
\end{tabular}
\end{adjustbox}
\vspace{-0in}
\caption{The decrease in accuracy for \sys/RQE compared to \sys.}
\vspace{-0in}
\label{tab:acc_ablation}
\end{table}

For SE, storing the sum values $\sum_z b_{zj}$ in Eq.~\eqref{eq:hoq_matmul} for KV to avoid redundant computation only requires 2.2\%-2.7\% of GPU memory capacity in the experiment.
For RQE, storing the FP16 values for the last block of $V$ only requires 0.24\%-0.51\% of GPU memory capacity.



\subsection{Sensitivity Testing}


We investigated the impact of \sys's different quantization partition sizes on end-to-end performance for Llama-3.1 70B on A10G prefill instances with different datasets.
Table.~\ref{tab:acc_e2e_group_sizes} lists the increase in accuracy and the percentage increase in the average JCT across all requests for quantization partition sizes $\Pi$=32 and $\Pi$=64 compared to $\Pi$=128. While $\Pi$=32 achieves the highest accuracy with up to 1.53\% increase, it can increase the average JCT by up to 28\%, which means it is a trade-off between accuracy and end-to-end performance requirements.


\begin{table}[h]
\centering
\begin{adjustbox}{max width=\columnwidth}
\begin{tabular}{|c|c|c|c|c|c|c|c|c|}
\hline
& \multicolumn{2}{|c|}{IMDb} & \multicolumn{2}{|c|}{arXiv} & \multicolumn{2}{|c|}{Cocktail} & \multicolumn{2}{|c|}{HumanEval} \\
\hline
& Acc. & JCT & Acc. & JCT & Acc. & JCT & Acc. & JCT \\
\hline
$\Pi$=32 & 0.92-1.46\% & 17.1\% & 0.53-1.02\% & 25.2\% & 0.81-1.53\% & 28\% & 0.95-1.51\% & 13.8\% \\
\hline
$\Pi$=64 & 0.4-0.96\% & 5.9\% & 0.23-0.59\% & 8.3\% & 0.22-1.18\% & 9.2\% & 0.59-1.27\% & 5.1\% \\
\hline
\end{tabular}
\end{adjustbox}
\vspace{-0in}
\caption{The increase in accuracy and average JCT across requests compared to $\Pi$=128.}
\vspace{-0.1in}
\label{tab:acc_e2e_group_sizes}
\end{table}


\subsection{Scalability Testing}


\DEL{We conducted a scalibility testing with Llama-3.1 70B and the Cocktail to demonstrate the ability of \sys to mitigate network bottlenecks when the ratio $p$ of the number of prefill instances to the number of decode instances increases from 1 to 8.}

\begin{figure}[h]
    \centering
    \includegraphics[width=0.99\columnwidth]{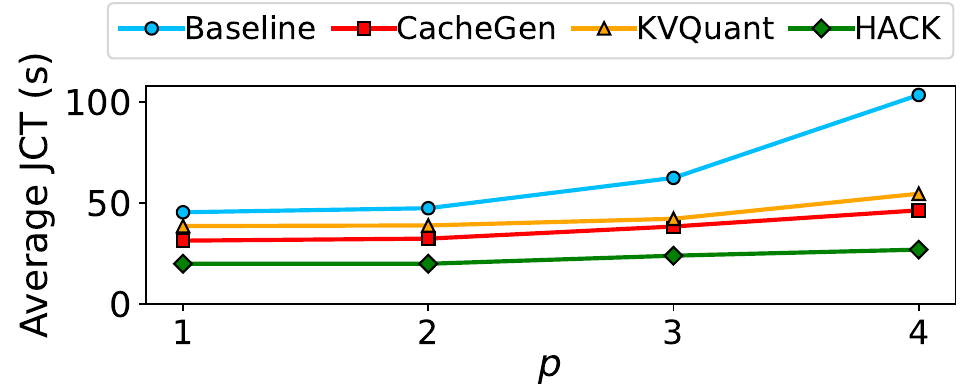}
    \vspace{-0.19in}
    \caption{Average JCT across requests with varying $p$.}
    \label{fig:e2e_scale}
    \vspace{-0.1in}
\end{figure}

We then test the ability of \sys to mitigate network bottlenecks using Llama-3.1 70B and Cocktail. We use $p$ to denote the ratio of the number of model replicas for prefill to the number of model replicas for decode.
The decode model ran on an A100 instance, and due to TP=4, only half of the A100 instance's GPU and network resources (i.e., 4 GPUs and a 200Gbps network) were allocated to serve it. The model replicas for prefill ran on A10G instances, and with TP=4 and PP=2, each prefill model required two A10G instances to serve it.
The RPS was set to $0.02p$.
Fig.~\ref{fig:e2e_scale} shows the average JCT across all requests when $p$ increases from 1 to 8. When $p$ increases from 1 to 8, the baseline's average JCT increases by 127\%, whereas CacheGen, KVQuant, and \sys only have an increase of 31-43\%. This demonstrates that CacheGen, KVQuant, and \sys effectively reduce the overhead caused by KV transmission at large scales.


\section{Limitations and Future Work}
Since one of our focuses is on mitigating the KV transmission bottleneck in disaggregated LLM inference, we employ 2-bit quantization to minimize KV transmission overhead. However, due to the low precision of 2-bit quantization, we must use small partition sizes to improve accuracy, which can increase JCT.
To address this, we plan to explore new quantization schemes that can reduce JCT and improve accuracy in disaggregated LLM inference in the future.
Additionally, since Triton only supports INT8 computation and incurs runtime overhead, we intend to implement \sys directly in CUDA to support INT4 computation and reduce execution overhead in out future work.


\section{Related Work}
\medskip
\noindent \textbf{Homomorphic Compression.}
The concept of homomorphic compression, where arithmetic operations are performed directly on compressed data, has been explored in the context of gradient aggregation~\cite{thc2024li}, enabling a parameter server to aggregate compressed gradients without decompressing them. However, this method is limited to addition operations and is unsuitable for the matrix multiplications required in the attention computation procedure.

\medskip
\noindent \textbf{Disaggregated LLM inference.} Disaggregated LLM inference.~\cite{distserve, splitwise, mooncake, memserve, strati2024dejavukvcachestreamingfast} has gained popularity as it separates the computation-intensive prefill and memory-intensive decode stages into different instances, improving resource utilization.
Two works~\cite{distserve, splitwise, strati2024dejavukvcachestreamingfast} focus on disaggregated LLM inference systems. Other research~\cite{mooncake, memserve} focuses on KV cache sharing, where KV data generated during the prefill stage is stored and shared across requests to reduce computation time. However, transmitting KV values from the prefill instance to the decode instance can become a bottleneck, a challenge that intensifies with KV data sharing. Our work addresses this issue by introducing homomorphic quantization, eliminating significant KV dequantization overhead.


\medskip
\noindent\textbf{KV Quantization.}
Many methods use quantization to compress KV~\cite{zipcache, kang2024gear, kivi, cachegen, kvquant} by reducing the high-bit representation FP16 to lower-bit representations in order to reduce the size of KV data. 
However, during attention computation, the quantized KV in the KV cache must first be dequantized to recover the original data, introducing significant decompression overhead. \sys, on the other hand, uses homomorphic quantization to avoid the KV dequantization overhead and performs attention computation on low-precision data to accelerate speed.


Recent work TurboAttention~\cite{turboattention} on arXiv, also performs attention computation directly on quantized KV data; however, it focuses solely on accelerating computation and mitigating KV cache bottlenecks. In contrast, our work not only accelerates computation and mitigates KV cache bottlenecks but also focuses on minimizing KV communication overhead in disaggregated LLM inference by leveraging 2-bit quantization.
Furthermore, TurboAttention chooses symmetric quantization to reduce computation overhead, which has a lower accuracy than asymmetric quantization~\cite{turboattention}. \sys uses asymmetric quantization with summation elimination to increase accuracy. In addition, \sys eliminates the requantization overhead and reduce quantization error for the last block of $V$ during the decode stage.

\medskip
\noindent\textbf{KV Eviction.}
Another solution for KV compression is KV eviction~\cite{h2o2023zhang, ge2023model, scissorhands2023liu, pyramidinfer, l2norm-kv, keyformer, dynamic-context-pruning, infinigen, zhang2024efficientsparseattentionneeds, jiang2024minference}, also known as KV pruning, which removes unimportant tokens' KV that have minimal impact on inference results, measured by attention scores.
Eviction-based methods and quantization-based methods are complementary: the former reduces the number of elements in the KV matrix, while the latter lowers the precision of those elements. These two approaches can be combined for enhanced effectiveness. While our work focuses on homomorphic quantization, we plan to explore how KV eviction can address the challenges in future work.

\DEL{
\noindent\textbf{TurboAttenion.}
Recent work TurboAttention~\cite{turboattention}, uploaded on arXiv, also performs attention computation directly on quantized KV data; however, it focuses solely on accelerating computation and mitigating KV cache bottlenecks. In contrast, our work not only accelerates computation and mitigates KV cache bottlenecks but also focuses on minimizing KV communication overhead in disaggregated LLM inference by leveraging 2-bit quantization.
Furthermore, as indicated in~\cref{sec:implementation}, TurboAttention uses symmetric quantization that decreases accuracy to eliminate approximation, while \sys uses asymmetric quantization with summation elimination to increase accuracy. \sys can also eliminate the requantization overhead and reduce quantization error for the last block of $V$ during the decode stage.
}


\section{Conclusion}
\sys is a novel quantization method for disaggregated LLM inference, which reduces the KV transmission overhead, computation time, and memory access latency for KV without introducing the expensive KV dequantization overhead. We integrate \sys into a widely adopted memory-efficient attention kernel, FlashAttention-2, and build our system on vLLM.
We conduct extensive experiments to demonstrate the efficiency of \sys, where \sys can reduce JCT by up to 70.9\% due to the improvement in communication, computation, memory access latency, and {the elimination of KV dequantization. }
This work does not raise any ethical issues.



\bibliographystyle{ACM-Reference-Format}
\bibliography{reference}


\begin{thebibliography}{52}


\ifx \showCODEN    \undefined \def \showCODEN     #1{\unskip}     \fi
\ifx \showDOI      \undefined \def \showDOI       #1{#1}\fi
\ifx \showISBNx    \undefined \def \showISBNx     #1{\unskip}     \fi
\ifx \showISBNxiii \undefined \def \showISBNxiii  #1{\unskip}     \fi
\ifx \showISSN     \undefined \def \showISSN      #1{\unskip}     \fi
\ifx \showLCCN     \undefined \def \showLCCN      #1{\unskip}     \fi
\ifx \shownote     \undefined \def \shownote      #1{#1}          \fi
\ifx \showarticletitle \undefined \def \showarticletitle #1{#1}   \fi
\ifx \showURL      \undefined \def \showURL       {\relax}        \fi
\providecommand\bibfield[2]{#2}
\providecommand\bibinfo[2]{#2}
\providecommand\natexlab[1]{#1}
\providecommand\showeprint[2][]{arXiv:#2}

\bibitem[\protect\citeauthoryear{??}{yi-}{2025}]%
        {yi-model}
 \bibinfo{year}{2025}\natexlab{}.
\newblock \bibinfo{title}{01-ai Model {Yi}}.
\newblock
  \bibinfo{howpublished}{\url{https://huggingface.co/01-ai/Yi-34B-200K}}.
  (\bibinfo{year}{2025}).
\newblock


\bibitem[\protect\citeauthoryear{??}{hkv}{2025}]%
        {hkvq-code}
 \bibinfo{year}{2025}\natexlab{}.
\newblock \bibinfo{title}{The code of {HKVQ}}.
\newblock \bibinfo{howpublished}{\url{https://anonymous.4open.science/r/HKVQ}}.
    (\bibinfo{year}{2025}).
\newblock


\bibitem[\protect\citeauthoryear{??}{fal}{2025}]%
        {falcon}
 \bibinfo{year}{2025}\natexlab{}.
\newblock \bibinfo{title}{Falcon-180B}.
\newblock
  \bibinfo{howpublished}{\url{https://huggingface.co/tiiuae/falcon-180B}}.
  (\bibinfo{year}{2025}).
\newblock


\bibitem[\protect\citeauthoryear{??}{sel}{2025}]%
        {self-attention}
 \bibinfo{year}{2025}\natexlab{}.
\newblock \bibinfo{title}{{GPT-4} explaining Self-Attention Mechanism}.
\newblock
  \bibinfo{howpublished}{\url{https://www.linkedin.com/pulse/gpt-4-explaining-self-attention-mechanism-fatos-ismali/}}.
    (\bibinfo{year}{2025}).
\newblock


\bibitem[\protect\citeauthoryear{??}{lla}{2025}]%
        {llama3.1}
 \bibinfo{year}{2025}\natexlab{}.
\newblock \bibinfo{title}{{Meta Llama-3.1}}.
\newblock \bibinfo{howpublished}{\url{https://llama.meta.com/}}.
  (\bibinfo{year}{2025}).
\newblock


\bibitem[\protect\citeauthoryear{??}{phi}{2025}]%
        {phi-3}
 \bibinfo{year}{2025}\natexlab{}.
\newblock \bibinfo{title}{{Microsoft Phi-3}}.
\newblock
  \bibinfo{howpublished}{\url{https://huggingface.co/microsoft/Phi-3-medium-128k-instruct}}.
    (\bibinfo{year}{2025}).
\newblock


\bibitem[\protect\citeauthoryear{??}{mis}{2025}]%
        {mistral-v0.3}
 \bibinfo{year}{2025}\natexlab{}.
\newblock \bibinfo{title}{{Mistral-v0.3}}.
\newblock
  \bibinfo{howpublished}{\url{https://huggingface.co/mistralai/Mistral-7B-Instruct-v0.3}}.
    (\bibinfo{year}{2025}).
\newblock


\bibitem[\protect\citeauthoryear{??}{ope}{2025}]%
        {open-compute}
 \bibinfo{year}{2025}\natexlab{}.
\newblock \bibinfo{title}{Open Compute Project}.
\newblock
  \bibinfo{howpublished}{\url{https://www.opencompute.org/documents/ocp-microscaling-formats-mx-v1-0-spec-final-pdf}}.
    (\bibinfo{year}{2025}).
\newblock


\bibitem[\protect\citeauthoryear{Adnan, Arunkumar, Jain, Nair, Soloveychik, and
  Kamath}{Adnan et~al\mbox{.}}{2024}]%
        {keyformer}
\bibfield{author}{\bibinfo{person}{Muhammad Adnan}, \bibinfo{person}{Akhil
  Arunkumar}, \bibinfo{person}{Gaurav Jain}, \bibinfo{person}{Prashant Nair},
  \bibinfo{person}{Ilya Soloveychik}, {and} \bibinfo{person}{Purushotham
  Kamath}.} \bibinfo{year}{2024}\natexlab{}.
\newblock \showarticletitle{Keyformer: KV Cache reduction through key tokens
  selection for Efficient Generative Inference}. In \bibinfo{booktitle}{{\em
  Proceedings of Machine Learning and Systems}},
  \bibfield{editor}{\bibinfo{person}{P.~Gibbons},
  \bibinfo{person}{G.~Pekhimenko}, {and} \bibinfo{person}{C.~De Sa}} (Eds.),
  Vol.~\bibinfo{volume}{6}. \bibinfo{pages}{114--127}.
\newblock
\showURL{%
\url{https://proceedings.mlsys.org/paper_files/paper/2024/file/48fecef47b19fe501d27d338b6d52582-Paper-Conference.pdf}}


\bibitem[\protect\citeauthoryear{Amazon Web~Services}{Amazon
  Web~Services}{2024}]%
        {aws-gpu-instances}
\bibfield{author}{\bibinfo{person}{Inc. Amazon Web~Services}.}
  \bibinfo{year}{2024}\natexlab{}.
\newblock \bibinfo{title}{Recommended {AWS GPU} Instances}.
\newblock
  \bibinfo{howpublished}{\url{https://docs.aws.amazon.com/dlami/latest/devguide/gpu.html}}.
    (\bibinfo{year}{2024}).
\newblock


\bibitem[\protect\citeauthoryear{Anagnostidis, Pavllo, Biggio, Noci, Lucchi,
  and Hofmann}{Anagnostidis et~al\mbox{.}}{2023}]%
        {dynamic-context-pruning}
\bibfield{author}{\bibinfo{person}{Sotiris Anagnostidis},
  \bibinfo{person}{Dario Pavllo}, \bibinfo{person}{Luca Biggio},
  \bibinfo{person}{Lorenzo Noci}, \bibinfo{person}{Aurelien Lucchi}, {and}
  \bibinfo{person}{Thomas Hofmann}.} \bibinfo{year}{2023}\natexlab{}.
\newblock \showarticletitle{Dynamic Context Pruning for Efficient and
  Interpretable Autoregressive Transformers}. In \bibinfo{booktitle}{{\em
  Advances in Neural Information Processing Systems}},
  \bibfield{editor}{\bibinfo{person}{A.~Oh}, \bibinfo{person}{T.~Naumann},
  \bibinfo{person}{A.~Globerson}, \bibinfo{person}{K.~Saenko},
  \bibinfo{person}{M.~Hardt}, {and} \bibinfo{person}{S.~Levine}} (Eds.),
  Vol.~\bibinfo{volume}{36}. \bibinfo{publisher}{Curran Associates, Inc.},
  \bibinfo{pages}{65202--65223}.
\newblock
\showURL{%
\url{https://proceedings.neurips.cc/paper_files/paper/2023/file/cdaac2a02c4fdcae77ba083b110efcc3-Paper-Conference.pdf}}


\bibitem[\protect\citeauthoryear{arXiv}{arXiv}{2025}]%
        {arxiv}
\bibfield{author}{\bibinfo{person}{arXiv}.} \bibinfo{year}{2025}\natexlab{}.
\newblock \bibinfo{title}{arXiv}.
\newblock \bibinfo{howpublished}{\url{https://arxiv.org}}.
  (\bibinfo{year}{2025}).
\newblock
\newblock
\shownote{Accessed: 2025-01-30.}


\bibitem[\protect\citeauthoryear{Chen, Tworek, Jun, Yuan, de~Oliveira~Pinto,
  Kaplan, Edwards, Burda, Joseph, Brockman, Ray, Puri, Krueger, Petrov, Khlaaf,
  Sastry, Mishkin, Chan, Gray, Ryder, Pavlov, Power, Kaiser, Bavarian, Winter,
  Tillet, Such, Cummings, Plappert, Chantzis, Barnes, Herbert-Voss, Guss,
  Nichol, Paino, Tezak, Tang, Babuschkin, Balaji, Jain, Saunders, Hesse, Carr,
  Leike, Achiam, Misra, Morikawa, Radford, Knight, Brundage, Murati, Mayer,
  Welinder, McGrew, Amodei, McCandlish, Sutskever, and Zaremba}{Chen
  et~al\mbox{.}}{2021}]%
        {humaneval}
\bibfield{author}{\bibinfo{person}{Mark Chen}, \bibinfo{person}{Jerry Tworek},
  \bibinfo{person}{Heewoo Jun}, \bibinfo{person}{Qiming Yuan},
  \bibinfo{person}{Henrique~Ponde de Oliveira~Pinto}, \bibinfo{person}{Jared
  Kaplan}, \bibinfo{person}{Harri Edwards}, \bibinfo{person}{Yuri Burda},
  \bibinfo{person}{Nicholas Joseph}, \bibinfo{person}{Greg Brockman},
  \bibinfo{person}{Alex Ray}, \bibinfo{person}{Raul Puri},
  \bibinfo{person}{Gretchen Krueger}, \bibinfo{person}{Michael Petrov},
  \bibinfo{person}{Heidy Khlaaf}, \bibinfo{person}{Girish Sastry},
  \bibinfo{person}{Pamela Mishkin}, \bibinfo{person}{Brooke Chan},
  \bibinfo{person}{Scott Gray}, \bibinfo{person}{Nick Ryder},
  \bibinfo{person}{Mikhail Pavlov}, \bibinfo{person}{Alethea Power},
  \bibinfo{person}{Lukasz Kaiser}, \bibinfo{person}{Mohammad Bavarian},
  \bibinfo{person}{Clemens Winter}, \bibinfo{person}{Philippe Tillet},
  \bibinfo{person}{Felipe~Petroski Such}, \bibinfo{person}{Dave Cummings},
  \bibinfo{person}{Matthias Plappert}, \bibinfo{person}{Fotios Chantzis},
  \bibinfo{person}{Elizabeth Barnes}, \bibinfo{person}{Ariel Herbert-Voss},
  \bibinfo{person}{William~Hebgen Guss}, \bibinfo{person}{Alex Nichol},
  \bibinfo{person}{Alex Paino}, \bibinfo{person}{Nikolas Tezak},
  \bibinfo{person}{Jie Tang}, \bibinfo{person}{Igor Babuschkin},
  \bibinfo{person}{Suchir Balaji}, \bibinfo{person}{Shantanu Jain},
  \bibinfo{person}{William Saunders}, \bibinfo{person}{Christopher Hesse},
  \bibinfo{person}{Andrew~N. Carr}, \bibinfo{person}{Jan Leike},
  \bibinfo{person}{Josh Achiam}, \bibinfo{person}{Vedant Misra},
  \bibinfo{person}{Evan Morikawa}, \bibinfo{person}{Alec Radford},
  \bibinfo{person}{Matthew Knight}, \bibinfo{person}{Miles Brundage},
  \bibinfo{person}{Mira Murati}, \bibinfo{person}{Katie Mayer},
  \bibinfo{person}{Peter Welinder}, \bibinfo{person}{Bob McGrew},
  \bibinfo{person}{Dario Amodei}, \bibinfo{person}{Sam McCandlish},
  \bibinfo{person}{Ilya Sutskever}, {and} \bibinfo{person}{Wojciech Zaremba}.}
  \bibinfo{year}{2021}\natexlab{}.
\newblock \bibinfo{title}{Evaluating Large Language Models Trained on Code}.
\newblock   (\bibinfo{year}{2021}).
\newblock
\showeprint[arxiv]{cs.LG/2107.03374}
\showURL{%
\url{https://arxiv.org/abs/2107.03374}}


\bibitem[\protect\citeauthoryear{Cloud}{Cloud}{2024}]%
        {tencentcloud-a100}
\bibfield{author}{\bibinfo{person}{Tencent Cloud}.}
  \bibinfo{year}{2024}\natexlab{}.
\newblock \bibinfo{title}{{Tencent Cloud} - {A100} Instances}.
\newblock
  \bibinfo{howpublished}{\url{https://www.tencentcloud.com/document/product/560/19701##GT4}}.
    (\bibinfo{year}{2024}).
\newblock


\bibitem[\protect\citeauthoryear{Cohan, Dernoncourt, Kim, Bui, Kim, Chang, and
  Goharian}{Cohan et~al\mbox{.}}{2018}]%
        {arxiv-summarization}
\bibfield{author}{\bibinfo{person}{Arman Cohan}, \bibinfo{person}{Franck
  Dernoncourt}, \bibinfo{person}{Doo~Soon Kim}, \bibinfo{person}{Trung Bui},
  \bibinfo{person}{Seokhwan Kim}, \bibinfo{person}{Walter Chang}, {and}
  \bibinfo{person}{Nazli Goharian}.} \bibinfo{year}{2018}\natexlab{}.
\newblock \bibinfo{title}{A Discourse-Aware Attention Model for Abstractive
  Summarization of Long Documents}.
\newblock   (\bibinfo{year}{2018}).
\newblock
\showeprint[arxiv]{cs.CL/1804.05685}
\showURL{%
\url{https://arxiv.org/abs/1804.05685}}


\bibitem[\protect\citeauthoryear{Corporation}{Corporation}{2025}]%
        {nccl}
\bibfield{author}{\bibinfo{person}{NVIDIA Corporation}.}
  \bibinfo{year}{2025}\natexlab{}.
\newblock \bibinfo{title}{NVIDIA NCCL Documentation}.
\newblock
  \bibinfo{howpublished}{\url{https://docs.nvidia.com/deeplearning/nccl/index.html}}.
    (\bibinfo{year}{2025}).
\newblock


\bibitem[\protect\citeauthoryear{Dai, Liu, Zhou, Pang, Ruan, Wang, Dong, Xu,
  and Wen}{Dai et~al\mbox{.}}{2024}]%
        {cocktailforir}
\bibfield{author}{\bibinfo{person}{Sunhao Dai}, \bibinfo{person}{Weihao Liu},
  \bibinfo{person}{Yuqi Zhou}, \bibinfo{person}{Liang Pang},
  \bibinfo{person}{Rongju Ruan}, \bibinfo{person}{Gang Wang},
  \bibinfo{person}{Zhenhua Dong}, \bibinfo{person}{Jun Xu}, {and}
  \bibinfo{person}{Ji-Rong Wen}.} \bibinfo{year}{2024}\natexlab{}.
\newblock \bibinfo{title}{Cocktail: A Comprehensive Information Retrieval
  Benchmark with LLM-Generated Documents Integration}.
\newblock   (\bibinfo{year}{2024}).
\newblock
\showeprint[arxiv]{cs.IR/2405.16546}
\showURL{%
\url{https://arxiv.org/abs/2405.16546}}


\bibitem[\protect\citeauthoryear{Dao}{Dao}{2023}]%
        {flashattn2}
\bibfield{author}{\bibinfo{person}{Tri Dao}.} \bibinfo{year}{2023}\natexlab{}.
\newblock \bibinfo{title}{FlashAttention-2: Faster Attention with Better
  Parallelism and Work Partitioning}.
\newblock   (\bibinfo{year}{2023}).
\newblock
\showeprint[arxiv]{cs.LG/2307.08691}
\showURL{%
\url{https://arxiv.org/abs/2307.08691}}


\bibitem[\protect\citeauthoryear{Devoto, Zhao, Scardapane, and
  Minervini}{Devoto et~al\mbox{.}}{2024}]%
        {l2norm-kv}
\bibfield{author}{\bibinfo{person}{Alessio Devoto}, \bibinfo{person}{Yu Zhao},
  \bibinfo{person}{Simone Scardapane}, {and} \bibinfo{person}{Pasquale
  Minervini}.} \bibinfo{year}{2024}\natexlab{}.
\newblock \showarticletitle{A Simple and Effective $L\_2$ Norm-Based Strategy
  for {KV} Cache Compression}. In \bibinfo{booktitle}{{\em Proceedings of the
  2024 Conference on Empirical Methods in Natural Language Processing}},
  \bibfield{editor}{\bibinfo{person}{Yaser Al-Onaizan}, \bibinfo{person}{Mohit
  Bansal}, {and} \bibinfo{person}{Yun-Nung Chen}} (Eds.).
  \bibinfo{publisher}{Association for Computational Linguistics},
  \bibinfo{address}{Miami, Florida, USA}, \bibinfo{pages}{18476--18499}.
\newblock
\showDOI{%
\url{https://doi.org/10.18653/v1/2024.emnlp-main.1027}}


\bibitem[\protect\citeauthoryear{Ge, Zhang, Liu, Zhang, Han, and Gao}{Ge
  et~al\mbox{.}}{2023}]%
        {ge2023model}
\bibfield{author}{\bibinfo{person}{Suyu Ge}, \bibinfo{person}{Yunan Zhang},
  \bibinfo{person}{Liyuan Liu}, \bibinfo{person}{Minjia Zhang},
  \bibinfo{person}{Jiawei Han}, {and} \bibinfo{person}{Jianfeng Gao}.}
  \bibinfo{year}{2023}\natexlab{}.
\newblock \showarticletitle{Model tells you what to discard: Adaptive kv cache
  compression for llms}.
\newblock \bibinfo{journal}{{\em arXiv preprint arXiv:2310.01801\/}}
  (\bibinfo{year}{2023}).
\newblock


\bibitem[\protect\citeauthoryear{Google}{Google}{2024}]%
        {gemini-1.5}
\bibfield{author}{\bibinfo{person}{Google}.} \bibinfo{year}{2024}\natexlab{}.
\newblock \bibinfo{title}{Gemini 1.5}.
\newblock \bibinfo{howpublished}{\url{https://gemini.google.com/app}}.
  (\bibinfo{year}{2024}).
\newblock


\bibitem[\protect\citeauthoryear{He, Zhang, Wu, Liu, Zhou, and Zhuang}{He
  et~al\mbox{.}}{2024}]%
        {zipcache}
\bibfield{author}{\bibinfo{person}{Yefei He}, \bibinfo{person}{Luoming Zhang},
  \bibinfo{person}{Weijia Wu}, \bibinfo{person}{Jing Liu},
  \bibinfo{person}{Hong Zhou}, {and} \bibinfo{person}{Bohan Zhuang}.}
  \bibinfo{year}{2024}\natexlab{}.
\newblock \bibinfo{title}{ZipCache: Accurate and Efficient KV Cache
  Quantization with Salient Token Identification}.
\newblock   (\bibinfo{year}{2024}).
\newblock
\showeprint[arxiv]{cs.LG/2405.14256}
\showURL{%
\url{https://arxiv.org/abs/2405.14256}}


\bibitem[\protect\citeauthoryear{Hooper, Kim, Mohammadzadeh, Mahoney, Shao,
  Keutzer, and Gholami}{Hooper et~al\mbox{.}}{2024}]%
        {kvquant}
\bibfield{author}{\bibinfo{person}{Coleman Richard~Charles Hooper},
  \bibinfo{person}{Sehoon Kim}, \bibinfo{person}{Hiva Mohammadzadeh},
  \bibinfo{person}{Michael~W. Mahoney}, \bibinfo{person}{Sophia Shao},
  \bibinfo{person}{Kurt Keutzer}, {and} \bibinfo{person}{Amir Gholami}.}
  \bibinfo{year}{2024}\natexlab{}.
\newblock \showarticletitle{{KVQ}uant: Towards 10 Million Context Length {LLM}
  Inference with {KV} Cache Quantization}. In \bibinfo{booktitle}{{\em The
  Thirty-eighth Annual Conference on Neural Information Processing Systems}}.
\newblock
\showURL{%
\url{https://openreview.net/forum?id=0LXotew9Du}}


\bibitem[\protect\citeauthoryear{Hu, Huang, Hu, Xu, Chen, Xie, Wang, Wang, Bao,
  Sun, and Shan}{Hu et~al\mbox{.}}{2024a}]%
        {memserve}
\bibfield{author}{\bibinfo{person}{Cunchen Hu}, \bibinfo{person}{Heyang Huang},
  \bibinfo{person}{Junhao Hu}, \bibinfo{person}{Jiang Xu},
  \bibinfo{person}{Xusheng Chen}, \bibinfo{person}{Tao Xie},
  \bibinfo{person}{Chenxi Wang}, \bibinfo{person}{Sa Wang},
  \bibinfo{person}{Yungang Bao}, \bibinfo{person}{Ninghui Sun}, {and}
  \bibinfo{person}{Yizhou Shan}.} \bibinfo{year}{2024}\natexlab{a}.
\newblock \bibinfo{title}{MemServe: Context Caching for Disaggregated LLM
  Serving with Elastic Memory Pool}.
\newblock   (\bibinfo{year}{2024}).
\newblock
\showeprint[arxiv]{cs.DC/2406.17565}
\showURL{%
\url{https://arxiv.org/abs/2406.17565}}


\bibitem[\protect\citeauthoryear{Hu, Huang, Xu, Chen, Xu, Chen, Feng, Wang,
  Wang, Bao, Sun, and Shan}{Hu et~al\mbox{.}}{2024b}]%
        {infer-without-infer}
\bibfield{author}{\bibinfo{person}{Cunchen Hu}, \bibinfo{person}{Heyang Huang},
  \bibinfo{person}{Liangliang Xu}, \bibinfo{person}{Xusheng Chen},
  \bibinfo{person}{Jiang Xu}, \bibinfo{person}{Shuang Chen},
  \bibinfo{person}{Hao Feng}, \bibinfo{person}{Chenxi Wang},
  \bibinfo{person}{Sa Wang}, \bibinfo{person}{Yungang Bao},
  \bibinfo{person}{Ninghui Sun}, {and} \bibinfo{person}{Yizhou Shan}.}
  \bibinfo{year}{2024}\natexlab{b}.
\newblock \bibinfo{title}{Inference without Interference: Disaggregate LLM
  Inference for Mixed Downstream Workloads}.
\newblock   (\bibinfo{year}{2024}).
\newblock
\showeprint[arxiv]{cs.DC/2401.11181}
\showURL{%
\url{https://arxiv.org/abs/2401.11181}}


\bibitem[\protect\citeauthoryear{IMDb}{IMDb}{2020}]%
        {imdb}
\bibfield{author}{\bibinfo{person}{IMDb}.} \bibinfo{year}{2020}\natexlab{}.
\newblock \bibinfo{title}{Genre Classification Dataset IMDb}.
\newblock
  \bibinfo{howpublished}{\url{https://www.kaggle.com/datasets/hijest/genre-classification-dataset-imdb}}.
    (\bibinfo{year}{2020}).
\newblock


\bibitem[\protect\citeauthoryear{Jiang, LI, Zhang, Wu, Luo, Ahn, Han, Abdi, Li,
  Lin, Yang, and Qiu}{Jiang et~al\mbox{.}}{2024}]%
        {jiang2024minference}
\bibfield{author}{\bibinfo{person}{Huiqiang Jiang}, \bibinfo{person}{YUCHENG
  LI}, \bibinfo{person}{Chengruidong Zhang}, \bibinfo{person}{Qianhui Wu},
  \bibinfo{person}{Xufang Luo}, \bibinfo{person}{Surin Ahn},
  \bibinfo{person}{Zhenhua Han}, \bibinfo{person}{Amir~H. Abdi},
  \bibinfo{person}{Dongsheng Li}, \bibinfo{person}{Chin-Yew Lin},
  \bibinfo{person}{Yuqing Yang}, {and} \bibinfo{person}{Lili Qiu}.}
  \bibinfo{year}{2024}\natexlab{}.
\newblock \showarticletitle{{MI}nference 1.0: Accelerating Pre-filling for
  Long-Context {LLM}s via Dynamic Sparse Attention}. In
  \bibinfo{booktitle}{{\em The Thirty-eighth Annual Conference on Neural
  Information Processing Systems}}.
\newblock
\showURL{%
\url{https://openreview.net/forum?id=fPBACAbqSN}}


\bibitem[\protect\citeauthoryear{Kang, Bharadwaj, Hensman, Krishna, Ruhle, and
  Rajmohan}{Kang et~al\mbox{.}}{2024a}]%
        {turboattention}
\bibfield{author}{\bibinfo{person}{Hao Kang}, \bibinfo{person}{Srikant
  Bharadwaj}, \bibinfo{person}{James Hensman}, \bibinfo{person}{Tushar
  Krishna}, \bibinfo{person}{Victor Ruhle}, {and} \bibinfo{person}{Saravan
  Rajmohan}.} \bibinfo{year}{2024}\natexlab{a}.
\newblock \bibinfo{title}{TurboAttention: Efficient Attention Approximation For
  High Throughputs LLMs}.
\newblock   (\bibinfo{year}{2024}).
\newblock
\showeprint[arxiv]{cs.LG/2412.08585}
\showURL{%
\url{https://arxiv.org/abs/2412.08585}}


\bibitem[\protect\citeauthoryear{Kang, Zhang, Kundu, Jeong, Liu, Krishna, and
  Zhao}{Kang et~al\mbox{.}}{2024b}]%
        {kang2024gear}
\bibfield{author}{\bibinfo{person}{Hao Kang}, \bibinfo{person}{Qingru Zhang},
  \bibinfo{person}{Souvik Kundu}, \bibinfo{person}{Geonhwa Jeong},
  \bibinfo{person}{Zaoxing Liu}, \bibinfo{person}{Tushar Krishna}, {and}
  \bibinfo{person}{Tuo Zhao}.} \bibinfo{year}{2024}\natexlab{b}.
\newblock \showarticletitle{{GEAR}: An Efficient {KV} Cache Compression Recipe
  for Near-Lossless Generative Inference of {LLM}}.
\newblock \bibinfo{journal}{{\em arXiv preprint arXiv:2403.05527\/}}
  (\bibinfo{year}{2024}).
\newblock
\showeprint[arxiv]{cs.LG/2403.05527}


\bibitem[\protect\citeauthoryear{Kiely}{Kiely}{2024}]%
        {a10g-inference}
\bibfield{author}{\bibinfo{person}{Philip Kiely}.}
  \bibinfo{year}{2024}\natexlab{}.
\newblock \bibinfo{title}{{NVIDIA A10 vs A10G for ML} model inference}.
\newblock
  \bibinfo{howpublished}{\url{https://www.baseten.co/blog/nvidia-a10-vs-a10g-for-ml-model-inference/}}.
    (\bibinfo{year}{2024}).
\newblock


\bibitem[\protect\citeauthoryear{Klauder}{Klauder}{1983}]%
        {squant}
\bibfield{author}{\bibinfo{person}{John~R. Klauder}.}
  \bibinfo{year}{1983}\natexlab{}.
\newblock \showarticletitle{Stochastic Quantization}. In
  \bibinfo{booktitle}{{\em Recent Developments in High-Energy Physics}},
  \bibfield{editor}{\bibinfo{person}{H.~Mitter} {and} \bibinfo{person}{C.~B.
  Lang}} (Eds.). \bibinfo{publisher}{Springer Vienna},
  \bibinfo{address}{Vienna}, \bibinfo{pages}{251--281}.
\newblock
\showISBNx{978-3-7091-7651-1}


\bibitem[\protect\citeauthoryear{Kwon, Li, Zhuang, Sheng, Zheng, Yu, Gonzalez,
  Zhang, and Stoica}{Kwon et~al\mbox{.}}{2023}]%
        {vllm2023kwon}
\bibfield{author}{\bibinfo{person}{Woosuk Kwon}, \bibinfo{person}{Zhuohan Li},
  \bibinfo{person}{Siyuan Zhuang}, \bibinfo{person}{Ying Sheng},
  \bibinfo{person}{Lianmin Zheng}, \bibinfo{person}{Cody~Hao Yu},
  \bibinfo{person}{Joseph Gonzalez}, \bibinfo{person}{Hao Zhang}, {and}
  \bibinfo{person}{Ion Stoica}.} \bibinfo{year}{2023}\natexlab{}.
\newblock \showarticletitle{Efficient Memory Management for Large Language
  Model Serving with PagedAttention}. In \bibinfo{booktitle}{{\em Proceedings
  of the 29th Symposium on Operating Systems Principles}} {\em
  (\bibinfo{series}{SOSP '23})}. \bibinfo{publisher}{Association for Computing
  Machinery}, \bibinfo{address}{New York, NY, USA}, \bibinfo{pages}{611–626}.
\newblock
\showISBNx{9798400702297}
\showDOI{%
\url{https://doi.org/10.1145/3600006.3613165}}


\bibitem[\protect\citeauthoryear{Lee, Lee, Seo, and Sim}{Lee
  et~al\mbox{.}}{2024}]%
        {infinigen}
\bibfield{author}{\bibinfo{person}{Wonbeom Lee}, \bibinfo{person}{Jungi Lee},
  \bibinfo{person}{Junghwan Seo}, {and} \bibinfo{person}{Jaewoong Sim}.}
  \bibinfo{year}{2024}\natexlab{}.
\newblock \showarticletitle{{InfiniGen}: Efficient Generative Inference of
  Large Language Models with Dynamic {KV} Cache Management}. In
  \bibinfo{booktitle}{{\em 18th USENIX Symposium on Operating Systems Design
  and Implementation (OSDI 24)}}. \bibinfo{publisher}{USENIX Association},
  \bibinfo{address}{Santa Clara, CA}, \bibinfo{pages}{155--172}.
\newblock
\showISBNx{978-1-939133-40-3}
\showURL{%
\url{https://www.usenix.org/conference/osdi24/presentation/lee}}


\bibitem[\protect\citeauthoryear{Li, Basat, Vargaftik, Lao, Xu, Mitzenmacher,
  and Yu}{Li et~al\mbox{.}}{2024}]%
        {thc2024li}
\bibfield{author}{\bibinfo{person}{Minghao Li}, \bibinfo{person}{Ran~Ben
  Basat}, \bibinfo{person}{Shay Vargaftik}, \bibinfo{person}{ChonLam Lao},
  \bibinfo{person}{Kevin Xu}, \bibinfo{person}{Michael Mitzenmacher}, {and}
  \bibinfo{person}{Minlan Yu}.} \bibinfo{year}{2024}\natexlab{}.
\newblock \showarticletitle{{THC}: Accelerating Distributed Deep Learning Using
  Tensor Homomorphic Compression}. In \bibinfo{booktitle}{{\em 21st USENIX
  Symposium on Networked Systems Design and Implementation (NSDI 24)}}.
  \bibinfo{publisher}{USENIX Association}, \bibinfo{address}{Santa Clara, CA},
  \bibinfo{pages}{1191--1211}.
\newblock
\showISBNx{978-1-939133-39-7}
\showURL{%
\url{https://www.usenix.org/conference/nsdi24/presentation/li-minghao}}


\bibitem[\protect\citeauthoryear{Lin}{Lin}{2004}]%
        {rouge-score}
\bibfield{author}{\bibinfo{person}{Chin-Yew Lin}.}
  \bibinfo{year}{2004}\natexlab{}.
\newblock \showarticletitle{Rouge: A package for automatic evaluation of
  summaries}. In \bibinfo{booktitle}{{\em Text summarization branches out}}.
  \bibinfo{pages}{74--81}.
\newblock


\bibitem[\protect\citeauthoryear{Liu, Li, Cheng, Ray, Huang, Zhang, Du, Yao,
  Lu, Ananthanarayanan, Maire, Hoffmann, Holtzman, and Jiang}{Liu
  et~al\mbox{.}}{2024a}]%
        {cachegen}
\bibfield{author}{\bibinfo{person}{Yuhan Liu}, \bibinfo{person}{Hanchen Li},
  \bibinfo{person}{Yihua Cheng}, \bibinfo{person}{Siddhant Ray},
  \bibinfo{person}{Yuyang Huang}, \bibinfo{person}{Qizheng Zhang},
  \bibinfo{person}{Kuntai Du}, \bibinfo{person}{Jiayi Yao},
  \bibinfo{person}{Shan Lu}, \bibinfo{person}{Ganesh Ananthanarayanan},
  \bibinfo{person}{Michael Maire}, \bibinfo{person}{Henry Hoffmann},
  \bibinfo{person}{Ari Holtzman}, {and} \bibinfo{person}{Junchen Jiang}.}
  \bibinfo{year}{2024}\natexlab{a}.
\newblock \showarticletitle{CacheGen: KV Cache Compression and Streaming for
  Fast Large Language Model Serving}. In \bibinfo{booktitle}{{\em Proceedings
  of the ACM SIGCOMM 2024 Conference}} {\em (\bibinfo{series}{ACM SIGCOMM
  '24})}. \bibinfo{publisher}{Association for Computing Machinery},
  \bibinfo{address}{New York, NY, USA}, \bibinfo{pages}{38–56}.
\newblock
\showISBNx{9798400706141}
\showDOI{%
\url{https://doi.org/10.1145/3651890.3672274}}


\bibitem[\protect\citeauthoryear{Liu, Desai, Liao, Wang, Xie, Xu, Kyrillidis,
  and Shrivastava}{Liu et~al\mbox{.}}{2023}]%
        {scissorhands2023liu}
\bibfield{author}{\bibinfo{person}{Zichang Liu}, \bibinfo{person}{Aditya
  Desai}, \bibinfo{person}{Fangshuo Liao}, \bibinfo{person}{Weitao Wang},
  \bibinfo{person}{Victor Xie}, \bibinfo{person}{Zhaozhuo Xu},
  \bibinfo{person}{Anastasios Kyrillidis}, {and} \bibinfo{person}{Anshumali
  Shrivastava}.} \bibinfo{year}{2023}\natexlab{}.
\newblock \showarticletitle{Scissorhands: Exploiting the Persistence of
  Importance Hypothesis for LLM KV Cache Compression at Test Time}. In
  \bibinfo{booktitle}{{\em Advances in Neural Information Processing Systems}},
  \bibfield{editor}{\bibinfo{person}{A.~Oh}, \bibinfo{person}{T.~Neumann},
  \bibinfo{person}{A.~Globerson}, \bibinfo{person}{K.~Saenko},
  \bibinfo{person}{M.~Hardt}, {and} \bibinfo{person}{S.~Levine}} (Eds.),
  Vol.~\bibinfo{volume}{36}. \bibinfo{publisher}{Curran Associates, Inc.},
  \bibinfo{pages}{52342--52364}.
\newblock
\showURL{%
\url{https://proceedings.neurips.cc/paper_files/paper/2023/file/a452a7c6c463e4ae8fbdc614c6e983e6-Paper-Conference.pdf}}


\bibitem[\protect\citeauthoryear{Liu, Yuan, Jin, Zhong, Xu, Braverman, Chen,
  and Hu}{Liu et~al\mbox{.}}{2024b}]%
        {kivi}
\bibfield{author}{\bibinfo{person}{Zirui Liu}, \bibinfo{person}{Jiayi Yuan},
  \bibinfo{person}{Hongye Jin}, \bibinfo{person}{Shaochen Zhong},
  \bibinfo{person}{Zhaozhuo Xu}, \bibinfo{person}{Vladimir Braverman},
  \bibinfo{person}{Beidi Chen}, {and} \bibinfo{person}{Xia Hu}.}
  \bibinfo{year}{2024}\natexlab{b}.
\newblock \showarticletitle{{KIVI}: A Tuning-Free Asymmetric 2bit Quantization
  for {KV} Cache}. In \bibinfo{booktitle}{{\em Proceedings of the 41st
  International Conference on Machine Learning}} {\em
  (\bibinfo{series}{Proceedings of Machine Learning Research})},
  \bibfield{editor}{\bibinfo{person}{Ruslan Salakhutdinov},
  \bibinfo{person}{Zico Kolter}, \bibinfo{person}{Katherine Heller},
  \bibinfo{person}{Adrian Weller}, \bibinfo{person}{Nuria Oliver},
  \bibinfo{person}{Jonathan Scarlett}, {and} \bibinfo{person}{Felix
  Berkenkamp}} (Eds.), Vol.~\bibinfo{volume}{235}. \bibinfo{publisher}{PMLR},
  \bibinfo{pages}{32332--32344}.
\newblock
\showURL{%
\url{https://proceedings.mlr.press/v235/liu24bz.html}}


\bibitem[\protect\citeauthoryear{OpenAI}{OpenAI}{2021}]%
        {triton}
\bibfield{author}{\bibinfo{person}{OpenAI}.} \bibinfo{year}{2021}\natexlab{}.
\newblock \bibinfo{title}{{Introducing Triton: Open-source GPU programming for
  neural networks}}.
\newblock \bibinfo{howpublished}{\url{https://openai.com/index/triton/}}.
  (\bibinfo{year}{2021}).
\newblock


\bibitem[\protect\citeauthoryear{Patel, Choukse, Zhang, Shah, Goiri, Maleki,
  and Bianchini}{Patel et~al\mbox{.}}{2024}]%
        {splitwise}
\bibfield{author}{\bibinfo{person}{Pratyush Patel}, \bibinfo{person}{Esha
  Choukse}, \bibinfo{person}{Chaojie Zhang}, \bibinfo{person}{Aashaka Shah},
  \bibinfo{person}{Íñigo Goiri}, \bibinfo{person}{Saeed Maleki}, {and}
  \bibinfo{person}{Ricardo Bianchini}.} \bibinfo{year}{2024}\natexlab{}.
\newblock \showarticletitle{Splitwise: Efficient Generative LLM Inference Using
  Phase Splitting}. In \bibinfo{booktitle}{{\em 2024 ACM/IEEE 51st Annual
  International Symposium on Computer Architecture (ISCA)}}.
  \bibinfo{pages}{118--132}.
\newblock
\showDOI{%
\url{https://doi.org/10.1109/ISCA59077.2024.00019}}


\bibitem[\protect\citeauthoryear{Qin, Li, He, Zhang, Wu, Zheng, and Xu}{Qin
  et~al\mbox{.}}{2024}]%
        {mooncake}
\bibfield{author}{\bibinfo{person}{Ruoyu Qin}, \bibinfo{person}{Zheming Li},
  \bibinfo{person}{Weiran He}, \bibinfo{person}{Mingxing Zhang},
  \bibinfo{person}{Yongwei Wu}, \bibinfo{person}{Weimin Zheng}, {and}
  \bibinfo{person}{Xinran Xu}.} \bibinfo{year}{2024}\natexlab{}.
\newblock \bibinfo{title}{Mooncake: A KVCache-centric Disaggregated
  Architecture for LLM Serving}.
\newblock   (\bibinfo{year}{2024}).
\newblock
\showeprint[arxiv]{cs.DC/2407.00079}
\showURL{%
\url{https://arxiv.org/abs/2407.00079}}


\bibitem[\protect\citeauthoryear{Strati, Mcallister, Phanishayee, Tarnawski,
  and Klimovic}{Strati et~al\mbox{.}}{2024}]%
        {strati2024dejavukvcachestreamingfast}
\bibfield{author}{\bibinfo{person}{Foteini Strati}, \bibinfo{person}{Sara
  Mcallister}, \bibinfo{person}{Amar Phanishayee}, \bibinfo{person}{Jakub
  Tarnawski}, {and} \bibinfo{person}{Ana Klimovic}.}
  \bibinfo{year}{2024}\natexlab{}.
\newblock \showarticletitle{DéjàVu: {KV}-cache Streaming for Fast,
  Fault-tolerant Generative {LLM} Serving}. In \bibinfo{booktitle}{{\em
  Proceedings of the 41st International Conference on Machine Learning}} {\em
  (\bibinfo{series}{Proceedings of Machine Learning Research})},
  \bibfield{editor}{\bibinfo{person}{Ruslan Salakhutdinov},
  \bibinfo{person}{Zico Kolter}, \bibinfo{person}{Katherine Heller},
  \bibinfo{person}{Adrian Weller}, \bibinfo{person}{Nuria Oliver},
  \bibinfo{person}{Jonathan Scarlett}, {and} \bibinfo{person}{Felix
  Berkenkamp}} (Eds.), Vol.~\bibinfo{volume}{235}. \bibinfo{publisher}{PMLR},
  \bibinfo{pages}{46745--46771}.
\newblock
\showURL{%
\url{https://proceedings.mlr.press/v235/strati24a.html}}


\bibitem[\protect\citeauthoryear{Tillet}{Tillet}{2020}]%
        {triton-flashattn}
\bibfield{author}{\bibinfo{person}{Philippe Tillet}.}
  \bibinfo{year}{2020}\natexlab{}.
\newblock \bibinfo{title}{Fused Attention in {Triton}}.
\newblock
  \bibinfo{howpublished}{\url{https://triton-lang.org/main/getting-started/tutorials/06-fused-attention.html}}.
    (\bibinfo{year}{2020}).
\newblock


\bibitem[\protect\citeauthoryear{vLLM Team}{vLLM Team}{2024}]%
        {vllm-fp8}
\bibfield{author}{\bibinfo{person}{vLLM Team}.}
  \bibinfo{year}{2024}\natexlab{}.
\newblock \bibinfo{title}{{vLLM FP8 for KV} Cache}.
\newblock
  \bibinfo{howpublished}{\url{https://docs.vllm.ai/en/v0.5.4/quantization/fp8_e4m3_kvcache.html}}.
    (\bibinfo{year}{2024}).
\newblock


\bibitem[\protect\citeauthoryear{Wang, Liu, Feng, Ding, and Ding}{Wang
  et~al\mbox{.}}{2024}]%
        {fp4}
\bibfield{author}{\bibinfo{person}{Jie Wang}, \bibinfo{person}{Huanxi Liu},
  \bibinfo{person}{Dawei Feng}, \bibinfo{person}{Jie Ding}, {and}
  \bibinfo{person}{Bo Ding}.} \bibinfo{year}{2024}\natexlab{}.
\newblock \showarticletitle{FP4-Quantization: Lossless 4bit Quantization for
  Large Language Models}. In \bibinfo{booktitle}{{\em 2024 IEEE International
  Conference on Joint Cloud Computing (JCC)}}. \bibinfo{pages}{61--67}.
\newblock
\showDOI{%
\url{https://doi.org/10.1109/JCC62314.2024.00017}}


\bibitem[\protect\citeauthoryear{Wu}{Wu}{2024}]%
        {string-similarity}
\bibfield{author}{\bibinfo{person}{Gang Wu}.} \bibinfo{year}{2024}\natexlab{}.
\newblock \bibinfo{title}{String Similarity Metrics – Edit Distance}.
\newblock
  \bibinfo{howpublished}{\url{https://www.baeldung.com/cs/string-similarity-edit-distance}}.
    (\bibinfo{year}{2024}).
\newblock


\bibitem[\protect\citeauthoryear{Xia, Zheng, Wu, Chen, Yao, Youn, Bakhtiari,
  Wyatt, Zhuang, Zhou, Ruwase, He, and Song}{Xia et~al\mbox{.}}{2024}]%
        {fp6}
\bibfield{author}{\bibinfo{person}{Haojun Xia}, \bibinfo{person}{Zhen Zheng},
  \bibinfo{person}{Xiaoxia Wu}, \bibinfo{person}{Shiyang Chen},
  \bibinfo{person}{Zhewei Yao}, \bibinfo{person}{Stephen Youn},
  \bibinfo{person}{Arash Bakhtiari}, \bibinfo{person}{Michael Wyatt},
  \bibinfo{person}{Donglin Zhuang}, \bibinfo{person}{Zhongzhu Zhou},
  \bibinfo{person}{Olatunji Ruwase}, \bibinfo{person}{Yuxiong He}, {and}
  \bibinfo{person}{Shuaiwen~Leon Song}.} \bibinfo{year}{2024}\natexlab{}.
\newblock \bibinfo{title}{FP6-LLM: Efficiently Serving Large Language Models
  Through FP6-Centric Algorithm-System Co-Design}.
\newblock   (\bibinfo{year}{2024}).
\newblock
\showeprint[arxiv]{cs.LG/2401.14112}
\showURL{%
\url{https://arxiv.org/abs/2401.14112}}


\bibitem[\protect\citeauthoryear{Yang, Han, Gao, Hu, Zhang, and Zhao}{Yang
  et~al\mbox{.}}{2024}]%
        {pyramidinfer}
\bibfield{author}{\bibinfo{person}{Dongjie Yang}, \bibinfo{person}{XiaoDong
  Han}, \bibinfo{person}{Yan Gao}, \bibinfo{person}{Yao Hu},
  \bibinfo{person}{Shilin Zhang}, {and} \bibinfo{person}{Hai Zhao}.}
  \bibinfo{year}{2024}\natexlab{}.
\newblock \bibinfo{title}{PyramidInfer: Pyramid KV Cache Compression for
  High-throughput LLM Inference}.
\newblock   (\bibinfo{year}{2024}).
\newblock
\showeprint[arxiv]{cs.CL/2405.12532}
\showURL{%
\url{https://arxiv.org/abs/2405.12532}}


\bibitem[\protect\citeauthoryear{Zhang, Zou, Luo, Tang, Luo, Li, and Li}{Zhang
  et~al\mbox{.}}{2024b}]%
        {zhang2024efficientsparseattentionneeds}
\bibfield{author}{\bibinfo{person}{Chaoran Zhang}, \bibinfo{person}{Lixin Zou},
  \bibinfo{person}{Dan Luo}, \bibinfo{person}{Min Tang},
  \bibinfo{person}{Xiangyang Luo}, \bibinfo{person}{Zihao Li}, {and}
  \bibinfo{person}{Chenliang Li}.} \bibinfo{year}{2024}\natexlab{b}.
\newblock \bibinfo{title}{Efficient Sparse Attention needs Adaptive Token
  Release}.
\newblock   (\bibinfo{year}{2024}).
\newblock
\showeprint[arxiv]{cs.CL/2407.02328}
\showURL{%
\url{https://arxiv.org/abs/2407.02328}}


\bibitem[\protect\citeauthoryear{Zhang, Li, Li, Xia, Yang, Luo, Wang, Chen,
  Liu, and Yang}{Zhang et~al\mbox{.}}{2024a}]%
        {zhang2024hierarchicalcontextpruningoptimizing}
\bibfield{author}{\bibinfo{person}{Lei Zhang}, \bibinfo{person}{Yunshui Li},
  \bibinfo{person}{Jiaming Li}, \bibinfo{person}{Xiaobo Xia},
  \bibinfo{person}{Jiaxi Yang}, \bibinfo{person}{Run Luo},
  \bibinfo{person}{Minzheng Wang}, \bibinfo{person}{Longze Chen},
  \bibinfo{person}{Junhao Liu}, {and} \bibinfo{person}{Min Yang}.}
  \bibinfo{year}{2024}\natexlab{a}.
\newblock \bibinfo{title}{Hierarchical Context Pruning: Optimizing Real-World
  Code Completion with Repository-Level Pretrained Code LLMs}.
\newblock   (\bibinfo{year}{2024}).
\newblock
\showeprint[arxiv]{cs.CL/2406.18294}
\showURL{%
\url{https://arxiv.org/abs/2406.18294}}


\bibitem[\protect\citeauthoryear{Zhang, Sheng, Zhou, Chen, Zheng, Cai, Song,
  Tian, R\'{e}, Barrett, Wang, and Chen}{Zhang et~al\mbox{.}}{2023}]%
        {h2o2023zhang}
\bibfield{author}{\bibinfo{person}{Zhenyu Zhang}, \bibinfo{person}{Ying Sheng},
  \bibinfo{person}{Tianyi Zhou}, \bibinfo{person}{Tianlong Chen},
  \bibinfo{person}{Lianmin Zheng}, \bibinfo{person}{Ruisi Cai},
  \bibinfo{person}{Zhao Song}, \bibinfo{person}{Yuandong Tian},
  \bibinfo{person}{Christopher R\'{e}}, \bibinfo{person}{Clark Barrett},
  \bibinfo{person}{Zhangyang~"Atlas" Wang}, {and} \bibinfo{person}{Beidi
  Chen}.} \bibinfo{year}{2023}\natexlab{}.
\newblock \showarticletitle{H2O: Heavy-Hitter Oracle for Efficient Generative
  Inference of Large Language Models}. In \bibinfo{booktitle}{{\em Advances in
  Neural Information Processing Systems}},
  \bibfield{editor}{\bibinfo{person}{A.~Oh}, \bibinfo{person}{T.~Neumann},
  \bibinfo{person}{A.~Globerson}, \bibinfo{person}{K.~Saenko},
  \bibinfo{person}{M.~Hardt}, {and} \bibinfo{person}{S.~Levine}} (Eds.),
  Vol.~\bibinfo{volume}{36}. \bibinfo{publisher}{Curran Associates, Inc.},
  \bibinfo{pages}{34661--34710}.
\newblock
\showURL{%
\url{https://proceedings.neurips.cc/paper_files/paper/2023/file/6ceefa7b15572587b78ecfcebb2827f8-Paper-Conference.pdf}}


\bibitem[\protect\citeauthoryear{Zhong, Liu, Chen, Hu, Zhu, Liu, Jin, and
  Zhang}{Zhong et~al\mbox{.}}{2024}]%
        {distserve}
\bibfield{author}{\bibinfo{person}{Yinmin Zhong}, \bibinfo{person}{Shengyu
  Liu}, \bibinfo{person}{Junda Chen}, \bibinfo{person}{Jianbo Hu},
  \bibinfo{person}{Yibo Zhu}, \bibinfo{person}{Xuanzhe Liu},
  \bibinfo{person}{Xin Jin}, {and} \bibinfo{person}{Hao Zhang}.}
  \bibinfo{year}{2024}\natexlab{}.
\newblock \showarticletitle{{DistServe}: Disaggregating Prefill and Decoding
  for Goodput-optimized Large Language Model Serving}. In
  \bibinfo{booktitle}{{\em 18th USENIX Symposium on Operating Systems Design
  and Implementation (OSDI 24)}}. \bibinfo{publisher}{USENIX Association},
  \bibinfo{address}{Santa Clara, CA}, \bibinfo{pages}{193--210}.
\newblock
\showISBNx{978-1-939133-40-3}
\showURL{%
\url{https://www.usenix.org/conference/osdi24/presentation/zhong-yinmin}}


\end{thebibliography}

\end{document}